\documentclass[aps,prd,superscriptaddress,groupedaddress,nofootinbib,nobibnotes]{revtex4}

\usepackage{graphicx}      
                                       
\usepackage{dcolumn}
\usepackage{bm}
\usepackage{amssymb}
\usepackage{epstopdf}
\usepackage{amsmath}
\usepackage{amsfonts}
\usepackage{amsthm}
\usepackage{color}
\usepackage{mathrsfs}
\usepackage{hyperref}
\usepackage{dsfont}
\usepackage{empheq}
\usepackage{float}
\usepackage[title]{appendix}
\usepackage[]{breqn}

\usepackage{scalerel}
\usepackage{tikz}
\usetikzlibrary{svg.path}
\definecolor{orcidlogocol}{HTML}{A6CE39}
\tikzset{
  orcidlogo/.pic={
    \fill[orcidlogocol] svg{M256,128c0,70.7-57.3,128-128,128C57.3,256,0,198.7,0,128C0,57.3,57.3,0,128,0C198.7,0,256,57.3,256,128z};
    \fill[white] svg{M86.3,186.2H70.9V79.1h15.4v48.4V186.2z}
                 svg{M108.9,79.1h41.6c39.6,0,57,28.3,57,53.6c0,27.5-21.5,53.6-56.8,53.6h-41.8V79.1z M124.3,172.4h24.5c34.9,0,42.9-26.5,42.9-39.7c0-21.5-13.7-39.7-43.7-39.7h-23.7V172.4z}
                 svg{M88.7,56.8c0,5.5-4.5,10.1-10.1,10.1c-5.6,0-10.1-4.6-10.1-10.1c0-5.6,4.5-10.1,10.1-10.1C84.2,46.7,88.7,51.3,88.7,56.8z};
  }
}
\usepackage{hyperref}  
\hypersetup{colorlinks}

\newcommand\orcidlink[1]{\href{https://orcid.org/#1}{\mbox{\scalerel*{
\begin{tikzpicture}[yscale=-1,transform shape]
\pic{orcidlogo};
\end{tikzpicture}
}{|}}}}

\usepackage{blindtext}
\begingroup\makeatletter
\catcode`,=\active
\global\let\breqn@comma,
\protected\gdef,{\ifmmode\expandafter\breqn@comma\else\expandafter\active@comma\fi}
\endgroup

\newcommand{\be}{\begin{equation}}
\newcommand{\ee}{\end{equation}}
\newcommand{\ba}{\begin{eqnarray}}
\newcommand{\ea}{\end{eqnarray}}

\newcommand{\barr}{\begin{array}}
\newcommand{\earr}{\end{array}}

\newcommand\lsim{\mathrel{\rlap{\lower4pt\hbox{\hskip1pt$\sim$}}
        \raise1pt\hbox{$<$}}}
\newcommand\gsim{\mathrel{\rlap{\lower4pt\hbox{\hskip1pt$\sim$}}
        \raise1pt\hbox{$>$}}}

\begin{document}

\title{Constraining Cosmological Parameters with Needlet Internal Linear Combination Maps I: Analytic Power Spectrum Formalism}
\date{\today}

\author{Kristen M.~Surrao\,\orcidlink{0000-0002-7611-6179}}
\email{k.surrao@columbia.edu}
\affiliation{Department of Physics, Columbia University, New York, NY 10027, USA}

\author{J.~Colin Hill\,\orcidlink{0000-0002-9539-0835}} 
\email{jch2200@columbia.edu}
\affiliation{Department of Physics, Columbia University, New York, NY 10027, USA}

\begin{abstract}
    The internal linear combination (ILC) method is a popular approach for constructing component-separated maps in cosmic microwave background (CMB) analyses. It optimally combines observed maps at different frequencies to produce an unbiased minimum-variance map of a component. When performed in harmonic space, it is straightforward to analytically compute the contributions of individual sky components to the power spectrum of the resulting ILC map. ILC can also be performed on a basis of needlets, spherical wavelets that have compact support in both pixel and harmonic space, capturing both scale-dependent and spatially varying information. However, an analytic understanding of the power spectra of needlet ILC (NILC) component-separated maps, as needed to enable their use in cosmological parameter inference, has remained an outstanding problem. In this paper, we derive the first analytic expression for the power spectra of NILC maps, as well as an expression for the cross-spectrum of a NILC map with an arbitrary second map, in terms of contributions from individual sky components. We validate our result using simulations, finding that it is exact. These results contain useful insights: we explicitly see how NILC power spectra contain information from contaminant fields beyond the two-point level, and we obtain a formalism with which to parameterize NILC power spectra. However, because this parameter dependence is complicated by correlations and higher-point functions of the component maps and weight maps, we find that it is intractable to perform parameter inference using these analytic expressions. Instead, numerical techniques are needed to estimate parameters using NILC maps --- we explore the use of likelihood-free inference with neural posterior estimation in a companion paper. Our code to produce the results in this paper is available in \verb|NILC-PS-Model|.\footnote{\url{https://github.com/kmsurrao/NILC-PS-Model}}
\end{abstract}

\maketitle

\section{Introduction}

In cosmic microwave background (CMB) data analyses, it is often useful to produce a best-estimate map of a specific signal in the sky \cite{Planck2018, Coulton2023ACT, Bleem2022SPT}. Several ``component-separation" techniques have been developed for this purpose.\footnote{\url{https://lambda.gsfc.nasa.gov/toolbox/comp_separation.html}} One class of component separation methods works by subtracting foreground templates, either in the map or spherical harmonic domain. Examples include Spectral Estimation via Expectation Maximization (SEVEM) \cite{2003MartinezGonzalez, 2008Leach, 2012FernandezCobos}, which constructs foreground templates by map-level differencing of low and high frequency channels, and Spectral Matching Independent Component Analysis (SMICA) \cite{2003Delabrouille, 2008Cardoso}, which works by comparing the data to a spectral domain model and fitting a minimal set of basis parameters without explicitly separating a map into components. Other techniques, such as Commander \cite{2006Eriksen, 2008Eriksen}, fit parametric models of the sky in each pixel and involve Monte Carlo methods.

Internal linear combination (ILC) \cite{Bennett2003, Tegmark2003, Eriksen2004} is a semi-blind approach that estimates the signal of interest by finding the minimum-variance linear combination of the observed frequency maps that satisfies the constraint of unit response to the signal of interest. It can be performed on several domains, including real/pixel space and harmonic space. Notably, it is often applied on a basis of needlets in what is known as needlet ILC, or NILC \cite{Delabrouille2009}. Needlets are a basis of spherical wavelets that have compact support in both pixel and harmonic space, allowing weights to vary as functions of both angular scale and position. Needlet ILC has been used to construct high-resolution maps of individual components using \emph{Planck}, the Atacama Cosmology Telescope (ACT), and the South Pole Telescope (SPT) data (e.g., Refs.~\cite{McCarthy:2023hpa, Chandran:2023akr, ACT:2023wcq, SPT-SZ:2021gsa, Remazeilles:2012pn}) and has also been used for simulation analysis of future experiments such as the Simons Observatory, CMB-S4, and \emph{LiteBIRD}, e.g., Refs.~\cite{Wolz:2023lzb, Carones:2022xzs, Abazajian:2019eic}.

While the use of NILC has been popular at the map level, it is also useful to study the power spectra of the resulting NILC maps. For example, several previous studies have computed and interpreted the power spectra of the \emph{Planck} NILC Compton-$y$ map \cite{Planck2015ymap, Bolliet2018, Rotti:2020rdl, Chandran:2023akr, McCarthy:2023hpa}. However, there have been far fewer attempts at using the power spectra of NILC component-separated maps in cosmological parameter inference. One notable attempt was made in the \emph{Planck} 2015 component separation analysis, which tried to perform inference using component-separated maps from various techniques, including NILC \cite{Planck2015compsep}. To estimate residual foreground contributions to the NILC power spectrum, they passed full focal plane 8 (FFP8) simulated foreground maps consisting of synchrotron, free-free, dust, CO, thermal SZ, and cosmic infrared background (CIB) emission through the NILC pipeline.  The resulting power spectrum was treated as a template in the NILC CMB map power spectrum, and was then marginalized over via a single amplitude parameter at the power spectrum level in the parameter inference process. The \emph{Planck} team found systematic uncertainties related to the foreground propagation at the $1$ -- $2 \sigma$ level due to this method of foreground modeling (by comparing with the best-fit $\Lambda$CDM model from the $\emph{Planck}$ 2015 multi-frequency power spectrum likelihood) and thus concluded that component-separated maps should not be used for robust cosmological parameter inference at that time. The single-parameter extragalactic foreground template was used because the propagation of foregrounds to NILC maps was thought to be too complicated to model. The reasoning is as follows: when ILC is performed in harmonic space, it is straightforward to compute the power spectrum of the resulting ILC map analytically since the weights are only $\ell$-dependent and not $m$-dependent. However, this is not the case for NILC, where the weights have nontrivial dependence on both $\ell$ and $m$ due to their spatially varying nature.  Thus, the power spectrum of the resulting NILC map is not a simple linear combination of the auto- and cross-power spectra of the sky components at each multipole.

In this paper, we provide the first analytic expression for the power spectra of NILC maps by treating the NILC weight maps as effective ``masks'' and building on standard techniques used in mask mode-coupling analyses. Our goal is to better model foreground propagation into NILC maps, so that it can be treated in a more rigorous way when doing parameter inference, hopefully resolving some of the systematic uncertainties that were observed in Ref.~\cite{Planck2015compsep}. Since NILC efficiently captures non-Gaussian information with its spatially varying weights, this approach could lead to smaller parameter error bars than traditional multifrequency power spectrum template-fitting techniques (used in current state-of-the-art cosmological parameter measurements with \emph{Planck} \cite{Planck2018}, ACT \cite{Dunkley, ACT2020}, and SPT \cite{SPT}) since essentially all CMB foregrounds are non-Gaussian. In particular, using a NILC weighting scheme, one could better suppress non-Gaussian CMB foregrounds, thus improving inference of the CMB component and yielding smaller cosmological parameter error bars. 

The remainder of this paper is organized as follows. In Sec.~\ref{sec.component_separation_methods}, we review ILC variants, specifically real-space ILC, harmonic ILC (HILC), and needlet ILC (NILC). In Sec.~\ref{sec:contributions}, we derive an analytic expression for the auto- and cross- power spectra of NILC component-separated maps, with the final result given in Eq.~\eqref{eq.NILC_PS_final}, and in Sec.~\ref{sec:computational_implementation} we explicitly validate this result using simulations. Finally, in Sec.~\ref{sec.discussion} we discuss our results and their interpretation, as well as ongoing follow-up work.

\section{Internal Linear Combination}
\label{sec.component_separation_methods}

In this section, we review formulations of the ILC method in various domains. The overall goal is to produce an unbiased map of some component of interest, while minimizing the variance of that map. To enforce the constraint that the final map is an unbiased map of the signal, we require it to have unit response to that signal, using its spectral energy distribution (SED). In this paper we review standard mininum-variance ILC methods, though it is also possible to impose constraints that deproject contaminants using their known SEDs \cite{2009Chen, 2011Remazeilles}, moment deprojection \cite{2017Chluba, Remazeilles:2020rqw}, or tracers \cite{Kusiak:2023hrz}.

\subsection{Real-Space ILC}
In pixel space, we seek to estimate the signal of interest in some pixel $p$ via a weighted linear combination of the frequency maps in that pixel: $\hat{T}(p) = \sum_i w_i \Delta T_i(p)$, where $\Delta T_i(p)$ is the temperature fluctuation in the $i$th frequency map in pixel $p$ and $w_i$ is the associated weight. Then the problem of finding the weights can be expressed as follows: 

\be
\begin{aligned}
\min_{w^i} \quad & \sigma^2_{\hat{T}}=N^{-1}_{\mathrm{pix}} \sum_p (\hat{T}(p)-\langle T \rangle)^2\\
\textrm{s.t.} \quad & \sum_i w^i g_i = 1 \, ,
\label{eq.pixel_ILC}
\end{aligned}
\ee
where $N_{\mathrm{pix}}$ is the number of pixels in the region of interest, $\langle T \rangle$ is the average signal across pixels in that region, and $g_i$ is the spectral response of the component of interest at the $i$th frequency channel (a vector of ones for the CMB, assuming the maps are in CMB thermodynamic temperature units). The first line of Eq.~\eqref{eq.pixel_ILC} minimizes the variance of the ILC map, while the second line enforces the constraint of unit response to the signal of interest. The solution for the weights can be found using Lagrange multipliers \cite{Eriksen2004}:

\be
\label{eq.real_space_cov}
w^i = \frac{\sum_j g_j(\hat{R}^{-1})_{ij}}{ \sum_{km}g_k(\hat{R}^{-1})_{km}g_m} \qquad \text{with} \qquad \hat{R}_{ij}=N^{-1}_{\mathrm{pix}}\sum_p \Delta T_i(p) \Delta T_j(p) \, ,
\ee 
where $\hat{R}_{ij}$ is the empirical frequency-frequency covariance matrix of the observed maps on the region of interest. Real-space ILC can also be performed on various subregions of a map to account for spatially-varying foregrounds. In that case, the ILC procedure is performed separately on each of the defined subregions, minimizing the variance of the final reconstructed map in that region alone. For instance, in the Wilkinson Microwave Anisotropy Probe (WMAP) analyses, the sky was partitioned into 12 regions \cite{Eriksen2004}.


\subsection{Harmonic-Space ILC}
The ILC method can also be formulated in harmonic space, giving $\ell$-dependent weights:
\be
\label{eq.HILCweights}
w^i_{\ell} = \frac{\sum_j \left( \hat{R}_{\ell}^{-1} \right)_{ij} g_j }{ \sum_{km} \left( \hat{R}_{\ell}^{-1} \right)_{km} g_k g_m }  \qquad \text{with} \qquad \left(\hat{R}_{\ell} \right)_{ij}= \sum_{\ell' = \ell-\Delta \ell /2}^{\ell+\Delta \ell /2} \frac{2\ell'+1}{4\pi} C_{\ell'}^{ij} \, .
\ee 
The multipole bin width $\Delta \ell$ must be large enough to mitigate the ``ILC bias" that results from computing the covariances for ILC weights using a finite number of modes \cite{Delabrouille2009}. 

In the harmonic ILC approach, we construct the harmonic transform of the ILC map, $\hat{T}_{\ell m}$, via
\be
\hat{T}_{\ell m} = \sum_i w^i_{\ell} T^i_{\ell m} \, .
\ee
The ILC weights are given by Eq.~\eqref{eq.HILCweights}, where the matrix of auto- and cross-power spectra $C_{\ell}^{ij}$ is just the frequency-frequency covariance matrix at this multipole.

As previously mentioned, it is straightforward to write down the power spectrum of a harmonic ILC map since the weights are only $\ell$-dependent and not $m$-dependent. The power spectrum is given by $C_\ell^{\hat{T}\hat{T}} = w^i_\ell w^j_\ell C^{ij}_\ell$.

\subsection{Needlet-Frame ILC}
\label{sec:needlet_ILC}

To maximize the robustness and flexibility of the ILC procedure, it is frequently applied to CMB data on a needlet frame \cite{Delabrouille2009, Planck:2013compsep, Planck:2013ymap, Planck2015compsep, Planck2015ymap, Planck2018, Chandran:2023akr, Coulton2023ACT, McCarthy:2023cwg, McCarthy:2023hpa}. Needlets are a set of basis functions on the sphere that possess compact support in both the harmonic and pixel-space domains \cite{Narcowich2006, Marinucci_2007}.  Thus, one can obtain ILC weights that vary both as a function of scale (i.e., depending on multipole $\ell$) and as a function of position (i.e., depending on direction or spatial pixel $\hat{n}$), allowing us to apply an optimal weighting scheme to non-Gaussian and/or non-isotropic foregrounds.  Here we briefly summarize the NILC procedure \cite{Guilloux2008, Delabrouille2009}.

Consider a set of maps $T^i(\hat{n})$, where $i$ labels the frequency channel of each map in our data set. For a needlet filter $h^{(n)}_{\ell}$, labeled by the index $(n)$ which ranges from $1$ to $N_{\rm scales}$ (the total number of needlet filter scales), the NILC operations on this set of frequency maps consist of the following steps:
\begin{itemize}
\item Filter each frequency map in harmonic space with the needlet filter function $h^{(n)}_{\ell}$:
\begin{equation}
\label{eq.filterstep1}
T^i_{\ell m} \rightarrow T^{i,(n)}_{\ell m} \equiv T^i_{\ell m} h^{(n)}_{\ell}
\end{equation}

\item Define some local domains in pixel space, and compute the smoothed frequency-frequency covariance matrix on each domain. Specifically, let $\mathcal{D}^{(n)}_{\alpha}$ denote a real-space domain in frequency maps that have been filtered with needlet scale $(n)$, where $\alpha$ labels each domain on that map. The frequency-frequency covariance matrix is then computed as 
\begin{equation}
    \label{eq.NILCcov}
    (\hat{R}_\alpha^{(n)})_{ij} = N_{\mathrm{pix}}^{-1} \sum_{p \in \mathcal{D}^{(n)}_{\alpha}} \Delta T_i(p) \Delta T_j(p) \, ,
\end{equation}
where $N_{\mathrm{pix}}$ is the number of pixels in $\mathcal{D}^{(n)}_{\alpha}$. Note that this equation is nearly the same as Eq.~\eqref{eq.real_space_cov} for the real-space ILC frequency-frequency covariance matrix, except that this is done independently for each needlet filter scale. In practice, Eq.~\eqref{eq.NILCcov} is usually implemented by smoothing the product map $\Delta T_i \Delta T_j$ with a Gaussian kernel, with larger kernels used for lower-multipole needlet filter scales.

\item For each needlet filter scale and each frequency channel, determine a map of weights in pixel space $W^{i,(n)}(\hat{n})$ in an analogous fashion to Eq.~\eqref{eq.real_space_cov}. The weight maps are determined via the ILC algorithm (i.e., Eq.~\eqref{eq.real_space_cov}) performed on the local pixel space domains from the previous step. Multiply each filtered frequency map by the corresponding weight map:
\begin{equation}
\label{eq.ILCweightingstep}
T^{i,(n)}(\hat{n}) \rightarrow \tilde{T}^{i,(n)}(\hat{n}) \equiv T^{i,(n)}(\hat{n}) W^{i,(n)}(\hat{n})
\end{equation}

\item Add up these ILC-weighted maps to obtain a single ILC map at this needlet filter scale:
\begin{equation}
\label{eq.ILCadditionstep}
T^{{\rm NILC},(n)}(\hat{n}) = \sum_i \tilde{T}^{i,(n)}(\hat{n})
\end{equation}

\item Apply the filter $h^{(n)}_{\ell}$ again:
\begin{equation}
\label{eq.filterstep2}
T^{{\rm NILC},(n)}_{\ell m} \rightarrow T^{{\rm NILC},(n),(n)}_{\ell m} \equiv T^{{\rm NILC},(n)}_{\ell m} h_{\ell}^{(n)}
\end{equation}

\item Add up the results from all filter scales to obtain the final NILC map:
\begin{equation}
\label{eq.ILCfinal}
T^{\rm NILC}(\hat{n}) = \sum_{(n)} T^{{\rm NILC},(n),(n)} (\hat{n})
\end{equation}

\end{itemize}
In creating a NILC map in this manner, the signal of interest will propagate in an unbiased fashion to the final map, due to the ILC signal-preservation constraint and the NILC filter power-preservation constraint, $\sum_{(n)} \left( h_{\ell}^{(n)} \right)^2 = 1$ at each $\ell$. However, residual contaminant signals will generally propagate in a nontrivial way, which we derive in Section~\ref{sec:contributions}.

\section{Derivation of NILC Power Spectra}
\label{sec:contributions}

We seek to derive an analytic expression that captures all contributions to NILC map power spectra, allowing us to easily see how various contaminants propagate. We derive an equation for the final power spectrum of a NILC map in terms of needlet filter functions, spectral response vectors, and $n$-point functions of the component maps and weight maps. 

\subsection{Notation and Definitions}
\label{sec.notation}

\subsubsection{Power Spectrum}

Letting $T(\mathbf{\hat{n}})$ represent the map of the signal of interest and $W(\mathbf{\hat{n}})$ represent some weight map (i.e., a ``mask'' or ``window'' in common CMB analysis terminology), we express the weighted (or masked) map as $\tilde T(\mathbf{\hat{n}})\equiv W(\mathbf{\hat{n}})T(\mathbf{\hat{n}})$. Each of these quantities can be expanded in harmonic space as follows:
\begin{equation}
    \label{eq.def_TMT-tilde}
    T(\mathbf{\hat{n}}) = \sum_{\ell,m} a_{\ell m}Y_{\ell m}(\mathbf{\hat{n}}), \qquad W(\mathbf{\hat{n}}) = \sum_{\ell, m} w_{\ell m}Y_{\ell m}(\mathbf{\hat{n}}), \qquad \tilde{T}(\mathbf{\hat{n}}) = \sum_{\ell,m} \tilde{a}_{\ell m}Y_{\ell m}(\mathbf{\hat{n}}),
\end{equation}
where $Y_{\ell m}(\mathbf{\hat{n}})$ are spherical harmonics and $\sum_{\ell, m} \equiv \sum_{\ell=0}^{\infty} \sum_{m=-\ell}^{\ell}$. The angular auto- and cross-power spectra of the map and mask/weight map are defined as
\begin{align}
    \langle a_{\ell_1 m_1}a_{\ell_2m_2}\rangle &= (-1)^{m_1} \langle a^*_{\ell_1 -m_1}a_{\ell_2m_2}\rangle \equiv (-1)^{m_1} C_{\ell_1}^{aa} \delta^{\rm K}_{\ell_1,\ell_2} \delta^{\rm K}_{m_1,-m_2} \nonumber \\
    \langle a_{\ell_1 m_1}w_{\ell_2m_2}\rangle &= (-1)^{m_1} \langle a^*_{\ell_1 -m_1}w_{\ell_2m_2}\rangle \equiv (-1)^{m_1} C_{\ell_1}^{aw} \delta^{\rm K}_{\ell_1,\ell_2} \delta^{\rm K}_{m_1,-m_2} \nonumber \\
    \langle w_{\ell_1 m_1}w_{\ell_2m_2}\rangle &= (-1)^{m_1} \langle w^*_{\ell_1 -m_1}w_{\ell_2m_2}\rangle \equiv (-1)^{m_1} C_{\ell_1}^{ww} \delta^{\rm K}_{\ell_1,\ell_2} \delta^{\rm K}_{m_1,-m_2},
\end{align} 
where $\delta^{\rm K}$ is the Kronecker delta-function and we have assumed statistical isotropy of the maps and masks.

\subsubsection{Bispectrum}

The connected bispectrum, or the connected three-point function, consisting of two factors of the map and one factor of the mask is defined as \cite{Komatsu:2001rj, Fergusson:2008, fergusson2011optimal, bucher2016binned}
\begin{equation}
    \langle a_{\ell_1 m_1} a_{\ell_2 m_2} w_{\ell_3 m_3} \rangle_c \equiv  B^{\ell_1 \ell_2 \ell_3}_{m_1 m_2 m_3}[aaw] \equiv \mathcal{G}^{\ell_1 \ell_2 \ell_3}_{m_1 m_2 m_3} b_{\ell_1 \ell_2 \ell_3}^{aaw},
    \label{eq.reduced_bispectrum}
\end{equation} where $B^{\ell_1 \ell_2 \ell_3}_{m_1 m_2 m_3}[aaw]$ is the connected bispectrum and $b^{aaw}_{\ell_1 \ell_2 \ell_3}$ is the reduced bispectrum, which is symmetric under any permutation of the joint set $\{(\ell_1,a), (\ell_2,a), (\ell_3,w)\}$.  In Eq.~\eqref{eq.reduced_bispectrum}, $\mathcal{G}^{\ell_1 \ell_2 \ell_3}_{m_1 m_2 m_3}$ represents the Gaunt integral, which is symmetric under exchange of $(\ell,m)$ pairs and can be expanded in terms of Wigner $3j$ symbols:
\begin{equation}
    \mathcal{G}^{\ell_1 \ell_2 \ell_3}_{m_1 m_2 m_3} \equiv \int d\mathbf{\hat{n}} \, 
    Y_{\ell_1 m_1}(\mathbf{\hat{n}}) Y_{\ell_2 m_2}(\mathbf{\hat{n}}) Y_{\ell_3 m_3}(\mathbf{\hat{n}}) = \sqrt{\frac{(2\ell_1+1)(2\ell_2+1)(2\ell_3+1)}{4\pi}} \begin{pmatrix} \ell_1&\ell_2&\ell_3 \\ 0&0&0 \end{pmatrix} \begin{pmatrix} \ell_1&\ell_2&\ell_3 \\ m_1&m_2&m_3 \end{pmatrix}
    \label{eq.def_gaunt}
\end{equation}

\subsubsection{Trispectrum}

The connected trispectrum, or the connected four-point function, consisting of two factors of the map and two factors of the mask is defined as \cite{Regan:2010, Fergusson:2010}
\begin{equation}
    \label{eq.trispectrum_def}
    \langle a_{\ell_1m_1} a_{\ell_2m_2} w_{\ell_3m_3} w_{\ell_4m_4} \rangle_c \equiv \sum_{L=0}^{\infty} \sum_{M=-L}^{L} (-1)^M \mathcal{G}^{\ell_1 \ell_2 L}_{m_1 m_2 -M} \mathcal{G}^{\ell_3 \ell_4 L}_{m_3 m_4 M} t[aaww]^{\ell_1 \ell_2}_{\ell_3 \ell_4}(L) + \text{ 23 perms.},
\end{equation}
 where the permutations are taken over the joint set $\{(\ell_1,m_1,a), (\ell_2,m_2,a), (\ell_3,m_3,w), (\ell_4,m_4,w) \}$ and $t[aaww]^{\ell_1 \ell_2}_{\ell_3 \ell_4}(L)$ is the reduced (parity-even) trispectrum. For two different fields $a$ and $w$, the reduced trispectrum has the following symmetries: 
 \begin{equation}
    \label{eq.reduced_trispectrum_symmetries}
     t[aaww]^{\ell_1 \ell_2}_{\ell_3 \ell_4}(L) = t[aaww]^{\ell_2 \ell_1}_{\ell_3 \ell_4}(L) = t[aaww]^{\ell_1 \ell_2}_{\ell_4 \ell_3}(L) = t[wwaa]^{\ell_3 \ell_4}_{\ell_1 \ell_2}(L)  \, ,
 \end{equation}
giving a set of eight equal permutations of the reduced trispectrum.

 We also define the estimator 
 \begin{equation}
    \label{eq.rho_def}
     \hat{\rho}[awaw]^{\ell_1 \ell_3}_{\ell_2 \ell_4}(L) \equiv \sum_{m_1,m_2,m_3,m_4,M} (-1)^M \mathcal{G}_{m_1 m_3 -M}^{\ell_1 \ell_3 L} \mathcal{G}_{m_2 m_4 M}^{\ell_2 \ell_4 L} a_{\ell_1 m_1} a_{\ell_2 m_2} w_{\ell_3 m_3} w_{\ell_4 m_4} \,.
 \end{equation}
The expectation of $\hat{\rho}[awaw]^{\ell_1 \ell_3}_{\ell_2 \ell_4}(L)$ is then (from Eq.~\eqref{eq.trispectrum_def})
 \begin{eqnarray}
     \label{eq.rho_expectation}
     \langle \hat{\rho}[awaw]^{\ell_1 \ell_3}_{\ell_2 \ell_4}(L) \rangle &=& \sum_{m_1 m_2 m_3 m_4} \sum_{L' M M'} (-1)^{M+M'} \mathcal{G}_{m_1 m_3 -M}^{\ell_1 \ell_3 L} \mathcal{G}_{m_2 m_4 M}^{\ell_2 \ell_4 L} \\\nonumber
     &&\,\times\,\left[\mathcal{G}_{m_1 m_2 -M'}^{\ell_1 \ell_2 L'} \mathcal{G}_{m_3 m_4 M'}^{\ell_3 \ell_4 L'} t[aaww]^{\ell_1 \ell_2}_{\ell_3 \ell_4}(L') + \text{ 23 perms.} \right],
 \end{eqnarray}
 where the 23 additional permutations are taken over only the piece in brackets. This estimator is the first step to measuring the trispectrum from data before we have to decorrelate and normalize it, as explained in detail in Ref.~\citep{PhilcoxNpoint}. The decorrelation/normalization is necessary since $\langle{\hat\rho}^{\ell_1\ell_2}_{\ell_3\ell_4}(L)\rangle$ depends on $t^{\ell_1\ell_2}_{\ell_3\ell_4}(L')$ with $L\neq L'$. Such correlation stems from the fact that there are two distinct internal legs that can be used to parameterize a quadrilateral. Thus, the decorrelation makes the reduced trispectrum diagonal in $L$.

\subsubsection{Components, Frequencies, and Filter Scales}

We use indices $y$ and $z$ to denote different sky components, such as the CMB and tSZ effect. For example, $C_{\ell}^{yz}$ is the cross-spectrum of components $y$ and $z$. Indices $\hat{p}$ and $\hat{q}$ denote NILC maps with preserved components $p$ and $q$, respectively. For example, $C_{\ell}^{\hat{p}\hat{q}}$ denotes the cross-spectrum of a NILC map with preserved component $p$ and a NILC map with preserved component $q$. Frequencies are indexed with $i$ and $j$. For example, $g^y_i$ is the spectral response of component $y$ at the $i$th frequency. Needlet filter scales are indexed with $(n)$ and $(m)$. For example, $h_{\ell}^{(n)}$ is the needlet filter at scale $n$. This information is summarized in Table \ref{table:notation}.

\begin{table}[t]
    \setlength{\tabcolsep}{10pt}
    \renewcommand{\arraystretch}{1.3}
    \centering
    \begin{tabular}{|c|c|c|}
        \hline
         \textbf{Indices} & \textbf{Meaning}  & \textbf{Sample Quantities}\\
         \hline
         $y,z$ & Sky Components & $C_{\ell}^{yz}$ (cross-spectrum of components $y$ and $z$)\\
         \hline
         $\hat{p},\hat{q}$ & Preserved Components  & $C_{\ell}^{\hat{p}\hat{q}}$ (cross-spectrum of NILC map with preserved component $p$ \\
          & in NILC maps&and NILC map with preserved component $q$)\\
          \hline
         $i,j$ & Frequencies & $g^y_i$ (spectral response of component $y$ at the $i$th frequency)\\
         \hline
         $(n),(m)$ & Needlet Filter Scales & $h_{\ell}^{(n)}$ (needlet filter at scale $(n)$)\\
          \hline
    \end{tabular}
    \caption{Notation and meaning of indices used for NILC power spectrum modeling.}
    \label{table:notation}
\end{table}

Some examples of the notation are as follows: $C_{\ell}^{y, pi(n)}$ is the cross-spectrum of component $y$ and the weight map at frequency $i$ and scale $(n)$ for a NILC map with preserved component $p$. $b_{\ell_1 \ell_2 \ell_3}^{y,z,pi(n)}$ is the reduced bispectrum involving component $y$, component $z$, and the weight map at frequency $i$ and scale $(n)$ for a NILC map with preserved component $p$. $b_{\ell_1 \ell_2 \ell_3}^{y,pi(n), qj(m)}$ is the reduced bispectrum involving component $y$, the weight map at frequency $i$ and scale $(n)$ for a NILC map with preserved component $p$, and the weight map at frequency $j$ and scale $(m)$ for a NILC map with preserved component $q$. Finally, $\langle \hat{\rho}[y,pi(n),z,qj(m)]^{\ell_2 \ell_4}_{\ell_3 \ell_5}(\ell_1) \rangle$ is the estimator $\hat{\rho}$ involving one factor of component $y$, one factor of the weight map at frequency $i$ and scale $(n)$ for a NILC map with preserved component $p$, one factor of component $z$, and one factor of the weight map at frequency $j$ and scale $(m)$ for a NILC map with preserved component $q$.

\subsection{Derivation of Auto- and Cross-Power Spectra of NILC Maps}
In this section, we derive an analytic expression for all contributions to the power spectrum of a NILC map, as well as cross-spectra of such maps. Our starting point is Eq.~\eqref{eq.ILCfinal}, which gives the expression for a NILC map in pixel space. To obtain the spherical harmonic transform of this expression, we follow the steps in \S \ref{sec:needlet_ILC}, working backwards:
\begin{equation}
    T^{\mathrm{NILC}}_{\ell_1 m_1} = \sum_{(n)}  T^{\mathrm{NILC},(n),(n)}_{\ell_1 m_1} = \sum_{(n)} h^{(n)}_{\ell_1} T^{\mathrm{NILC},(n)}_{\ell_1 m_1} = \sum_i \sum_{(n)} h^{(n)}_{\ell_1} \tilde{T}^{i,(n)}_{\ell_1 m_1} = \sum_i \sum_{(n)} h^{(n)}_{\ell_1} \left[ T^{i,(n)} W^{i,(n)} \right]_{\ell_1 m_1} \,.
    \label{eq.Tell1m1firststep}
\end{equation}
We thus have an expression for the spherical harmonic transform of a NILC map in terms of needlet filters and the harmonic transforms of weighted frequency maps. The next step is to explicitly compute the harmonic transform of a weighted frequency map. Motivated by the MASTER formalism \cite{Hivon2002}, which gives a simple relationship for the power spectrum of a masked map in terms of the power spectrum of the unmasked map and the power spectrum of the mask, we express the spherical harmonic transform of a ``masked" map (here we treat the weight map as an effective ``mask") by doing the following:
\begin{align}
    [T^{i,(n)}W^{i,(n)}]_{\ell_1 m_1} &\equiv \int  d\mathbf{\hat{n}}\, T^{i,(n)}(\mathbf{\hat{n}})W^{i,(n)}(\mathbf{\hat{n}})Y^{*}_{\ell_1 m_1}(\mathbf{\hat{n}}) \nonumber \\
    & = \sum_{\ell_2m_2}T^{i,(n)}_{\ell_2m_2}\int d\mathbf{\hat{n}}\, Y_{\ell_2m_2}(\mathbf{\hat{n}})W^{i,(n)}(\mathbf{\hat{n}})Y^{*}_{\ell_1 m_1}(\mathbf{\hat{n}}) \nonumber \\
    &=\sum_{\ell_2 m_2} \sum_{\ell_4 m_4} (-1)^{m_1} \mathcal{G}^{\ell_1 \ell_2 \ell_4}_{-m_1 m_2 m_4} W^{i,(n)}_{\ell_4 m_4} T^{i,(n)}_{\ell_2 m_2} \nonumber \\
    & = \sum_{\ell_2 m_2} \sum_{\ell_4 m_4} (-1)^{m_1} \mathcal{G}^{\ell_1 \ell_2 \ell_4}_{-m_1 m_2 m_4} h^{(n)}_{\ell_2} W^{i,(n)}_{\ell_4 m_4} T^{i}_{\ell_2 m_2} \,.
    \label{eq.maskedmapcoeffs}
\end{align}
In going from the second to the third line above, we expanded the weight map $W^{i,(n)}(\mathbf{\hat{n}})$ (i.e., the weight map for the $i$th frequency channel at needlet filter scale $(n)$) in terms of spherical harmonic coefficients, $W^{i,(n)}_{\ell_4 m_4}$, and then used the definition of the Gaunt integral from Eq.~\eqref{eq.def_gaunt}. Using Eq.~\eqref{eq.maskedmapcoeffs} in Eq.~\eqref{eq.Tell1m1firststep}, we thus obtain
\begin{equation}
    \label{eq.NILC_map}
    T^{\mathrm{NILC}}_{\ell_1 m_1} = \sum_{i} \sum_{(n)} \sum_{\ell_2 m_2} \sum_{\ell_4 m_4} (-1)^{m_1} \mathcal{G}^{\ell_1 \ell_2 \ell_4}_{-m_1 m_2 m_4} h^{(n)}_{\ell_1} h^{(n)}_{\ell_2} W^{i,(n)}_{\ell_4 m_4} T^{i}_{\ell_2 m_2} \,.
\end{equation}
This result shows how the spherical harmonic transform of a NILC map can be simply expressed in terms of needlet filters, the input frequency maps, and weights maps determined via the NILC algorithm.

With the harmonic coefficients on hand, we can proceed to evaluate auto- and cross-power spectra of NILC maps.  To obtain the cross-spectrum of a NILC map with some other NILC map, we add an index denoting which component is preserved for the construction of each NILC map, since the preserved component dictates the weight maps. Thus, to denote a weight map at frequency $i$ and scale $(n)$ used to build a NILC map with some preserved component $p$, we use $W^{pi(n)}$. Then the cross-spectrum of a NILC map with preserved component $p$ and a NILC map with preserved component $q$ is 
\begin{align}
    \label{eq.NILC_PS_prelim}
    \langle C_{\ell_1}^{\hat{p} \hat{q}} \rangle &= \frac{1}{2\ell_1+1} \sum_{m_1=-\ell_1}^{\ell_1} \sum_{i,j} \sum_{(n),(m)} \sum_{\ell_2 m_2} \sum_{\ell_3 m_3} \sum_{\ell_4 m_4} \sum_{\ell_5 m_5}  h_{\ell_1}^{(n)} h_{\ell_1}^{(m)} h_{\ell_2}^{(n)} h_{\ell_3}^{(m)} \mathcal{G}^{\ell_1 \ell_2 \ell_4}_{-m_1 m_2 m_4} \mathcal{G}^{\ell_1 \ell_3 \ell_5}_{-m_1 m_3 m_5} \nonumber
    \\&\qquad \qquad \qquad \qquad \qquad \qquad \qquad \qquad \qquad \qquad \times \langle W^{pi(n)}_{\ell_4 m_4} T^{i}_{\ell_2 m_2} (W^{qj(m)}_{\ell_5 m_5})^* (T^{j}_{\ell_3 m_3})^* \rangle \,.
\end{align}
The above result expresses NILC power spectra in terms of needlet filters and a four-point correlation function of the input frequency maps and NILC-determined weight maps. If the weight maps and frequency maps were uncorrelated, we could rewrite the four-point function as the product of two two-point functions, where one two-point function involves the weight maps alone, and the other two-point function involves the frequency maps alone, i.e.,
\begin{equation*}
    \langle W^{pi(n)}_{\ell_4 m_4} T^{i}_{\ell_2 m_2} (W^{qj(m)}_{\ell_5 m_5})^* (T^{j}_{\ell_3 m_3})^* \rangle = \langle W^{pi(n)}_{\ell_4 m_4} (W^{qj(m)}_{\ell_5 m_5})^* \rangle \langle T^{i}_{\ell_2 m_2}  (T^{j}_{\ell_3 m_3})^* \rangle
\end{equation*}
We could then apply the standard MASTER result for relating the power spectrum of a masked map to an expression involving the product of the power spectrum of the mask and the power spectrum of the map. The problem here is that the weight maps and frequency maps are actually correlated, and thus the separation into two two-point functions is not possible. Instead, all of the Wick contractions contribute to the four-point function. The weight maps determined by the NILC algorithm are specifically constructed so as to suppress foreground contamination and instrumental noise, and will thus naturally be correlated with the frequency maps themselves. We thus require a slightly different approach.

Treating the weight map as an effective ``mask" and noting that the map at each frequency and the corresponding weight map are correlated by consequence of the ILC weight determination algorithm (the ILC weights are optimized to clean out contaminant fields and noise), we note that a subset of this problem is very similar to a problem that has been solved using the reMASTERed approach \cite{remastered}. The reMASTERed approach is a generalization of the MASTER approach \cite{Hivon2002} that provides correction terms to the MASTER result in the case that the map and mask fields are correlated. It states that the ensemble-averaged power spectrum of a masked map, $\langle \tilde{C}_{\ell_1} \rangle$, is given by 
\begin{align}
    \langle \tilde{C}_{\ell_1} \rangle &= \frac{1}{2\ell_1+1}\sum_{m_1=-\ell_1}^{\ell_1}\sum_{\ell_2 m_2}\sum_{\ell_3 m_3} \sum_{\ell_4 m_4} \sum_{\ell_5 m_5} \langle a_{\ell_2 m_2}a^{*}_{\ell_3 m_3} w_{\ell_4 m_4} w^{*}_{\ell_5 m_5} \rangle \mathcal{G}^{\ell_1 \ell_2 \ell_4}_{-m_1 m_2 m_4} \mathcal{G}^{\ell_1 \ell_3 \ell_5}_{-m_1 m_3 m_5} \nonumber
    \\&= \frac{1}{4\pi} \sum_{\ell_2,\ell_3} (2\ell_2+1)(2\ell_3+1) \begin{pmatrix} \ell_1&\ell_2&\ell_3 \\ 0&0&0 \end{pmatrix}^2 \left[  \langle C_{\ell_2}^{aa}\rangle
    \langle C_{\ell_3}^{ww}\rangle +  \langle C_{\ell_2}^{aw}\rangle  \langle C_{\ell_3}^{aw}\rangle + \frac{\langle w_{00} \rangle}{\sqrt{\pi}}  \langle b_{\ell_1 \ell_2 \ell_3}^{aaw} \rangle + \frac{\langle a_{00} \rangle}{\sqrt{\pi}}  \langle b_{\ell_1 \ell_2 \ell_3}^{waw} \rangle  \right] \notag \nonumber \\&\qquad + \frac{1}{2\ell_1+1} \sum_{\ell_2 \ell_3 \ell_4 \ell_5} \langle \hat{\rho}[awaw]^{\ell_2 \ell_4}_{\ell_3 \ell_5}(\ell_1) \rangle \, ,
\end{align}
where $a$ and $w$ denote the map and mask, respectively. In our case, the first difference is that we have two different temperature maps and two different weight maps. To make contact with Eq.~\eqref{eq.NILC_PS_prelim}, we revisit the reMASTERed derivation (Appendix A of Ref.~\cite{remastered}) and note that the result can be generalized as follows for the cross-power spectrum of two masked maps:
\begin{align}
    \label{eq.reMASTERed_general}
    \langle C_{\ell_1}^{\tilde{a} \tilde{a}'} \rangle &=  \frac{1}{2\ell_1+1}\sum_{m_1=-\ell_1}^{\ell_1}\sum_{\ell_2 m_2}\sum_{\ell_3 m_3} \sum_{\ell_4 m_4} \sum_{\ell_5 m_5} \langle a_{\ell_2 m_2}(a'_{\ell_3 m_3})^* w_{\ell_4 m_4} (w'_{\ell_5 m_5})^* \rangle \mathcal{G}^{\ell_1 \ell_2 \ell_4}_{-m_1 m_2 m_4} \mathcal{G}^{\ell_1 \ell_3 \ell_5}_{-m_1 m_3 m_5} \nonumber
    \\&= \frac{1}{4\pi} \sum_{\ell_2,\ell_3} (2\ell_2+1)(2\ell_3+1) \begin{pmatrix} \ell_1&\ell_2&\ell_3 \\ 0&0&0 \end{pmatrix}^2 \Bigg[  \langle C_{\ell_2}^{aa'}\rangle
    \langle C_{\ell_3}^{ww'}\rangle +  \langle C_{\ell_2}^{aw'}\rangle  \langle C_{\ell_3}^{a'w}\rangle   \nonumber
    \\&\qquad + \frac{\langle w_{00} \rangle}{2\sqrt{\pi}}  \langle b_{\ell_1 \ell_2 \ell_3}^{aa'w'} \rangle + \frac{\langle w'_{00} \rangle}{2\sqrt{\pi}}  \langle b_{\ell_1 \ell_2 \ell_3}^{aa'w} \rangle + \frac{\langle a_{00} \rangle}{2\sqrt{\pi}}  \langle b_{\ell_1 \ell_2 \ell_3}^{wa'w'} \rangle + \frac{\langle a'_{00} \rangle}{2\sqrt{\pi}}  \langle b_{\ell_1 \ell_2 \ell_3}^{waw'} \rangle \Bigg] 
    + \frac{1}{2\ell_1+1} \sum_{\ell_2 \ell_3 \ell_4 \ell_5} \langle \hat{\rho}[awa'w']^{\ell_2 \ell_4}_{\ell_3 \ell_5}(\ell_1) \rangle,
\end{align}
where $a$ and $a'$ denote the two maps, $w$ and $w'$ denote their respective masks, and $\tilde{a}$ and $\tilde{a}'$ denote the masked maps. The full derivation of this generalization is shown in Appendix \ref{app:reMASTERed_generalization}. Note that the first term alone on the right-hand side in the above result, involving $\langle C_{\ell_2}^{aa'}\rangle \langle C_{\ell_3}^{ww'}\rangle$, corresponds to using the standard MASTER result. Also, note that the first term does not assume isotropy of the mask/weight map; however, the other terms do assume isotropy of the mask/weight map in regions that are correlated with the signal map. Thus, the result is not exact for masks or weight maps that are correlated with the signal in an anisotropic manner, but is still likely to be a very good approximation.  In our simulation-based tests below, we do not see any evidence of inaccuracy arising from this subtlety.

We apply the result from Eq.~\eqref{eq.reMASTERed_general} to Eq.~\eqref{eq.NILC_PS_prelim}, noting that the latter two needlet filters' subscripts in Eq.~\eqref{eq.NILC_PS_prelim} must match those of the original maps they filtered (and the $\ell$ subscripts of these maps may have changed due to Kronecker delta functions and index relabeling in the reMASTERed derivation). Letting $a \rightarrow T^i$, $a' \rightarrow T^j$, $w \rightarrow pi(n)$, and $w' \rightarrow qj(m)$ (where $pi(n)$ is shorthand for $W^{pi(n)}$ and similarly for $qj(m)$), this yields
\begin{align}
    \label{eq.NILC_PS_insert_remastered}
    \langle C_{\ell_1}^{\hat{p} \hat{q}} \rangle &=  \sum_{i,j} \sum_{(n),(m)}   h_{\ell_1}^{(n)} h_{\ell_1}^{(m)} \Bigg( \frac{1}{4 \pi} \sum_{\ell_2,\ell_3} (2\ell_2+1)(2\ell_3+1) \begin{pmatrix} \ell_1&\ell_2&\ell_3 \\ 0&0&0 \end{pmatrix}^2 \nonumber
    \\&\times \Bigg[h_{\ell_2}^{(n)} h_{\ell_2}^{(m)}  \langle C_{\ell_2}^{T^i,T^j}\rangle
    \langle C_{\ell_3}^{pi(n),qj(m)}\rangle +  h_{\ell_2}^{(n)} h_{\ell_3}^{(m)} \langle C_{\ell_2}^{T^i, qj(m)}\rangle  \langle C_{\ell_3}^{T^j, pi(n)}\rangle + h_{\ell_1}^{(n)} h_{\ell_2}^{(m)} \frac{\langle w^{pi(n)}_{00} \rangle}{2\sqrt{\pi}}  \langle b_{\ell_1 \ell_2 \ell_3}^{T^i,T^j,qj(m)} \rangle \nonumber
    \\&\qquad + h_{\ell_1}^{(n)} h_{\ell_2}^{(m)} \frac{\langle w^{qj(m)}_{00} \rangle}{2\sqrt{\pi}}  \langle b_{\ell_1 \ell_2 \ell_3}^{T^i, T^j, pi(n)} \rangle + h_{0}^{(n)} h_{\ell_2}^{(m)} \frac{\langle T^i_{00} \rangle}{2\sqrt{\pi}}  \langle b_{\ell_1 \ell_2 \ell_3}^{pi(n),T^j,qj(m)} \rangle + h_{\ell_2}^{(n)} h_{0}^{(m)} \frac{\langle T^j_{00} \rangle}{2\sqrt{\pi}}  \langle b_{\ell_1 \ell_2 \ell_3}^{pi(n),T^i,qj(m)} \rangle \Bigg] \nonumber
    \\&+ \frac{1}{2\ell_1+1} \sum_{\ell_2 \ell_3 \ell_4 \ell_5} h_{\ell_2}^{(n)} h_{\ell_3}^{(m)}  \langle \hat{\rho}[T^i,pi(n),T^j,qj(m)]^{\ell_2 \ell_4}_{\ell_3 \ell_5}(\ell_1) \rangle \Bigg) \, .
\end{align}
Noting that the sky map at the $i$th frequency, $T^i$, consists of several sky components (labeled by $y$ with associated component map denoted $a^y(\hat{n})$) with spectral response $g_i^y \equiv g^y(\nu_i)$ at frequency $\nu_i$, we can write $T^i(\hat{n}) = \sum_y g_i^y a^y(\hat{n})$ and similarly, $T^j(\hat{n}) = \sum_z g_j^z a^z(\hat{n})$, giving 
\begin{equation}\label{eq.NILC_PS_final}
    \boxed{
    \begin{aligned}
    \langle C_{\ell_1}^{\hat{p} \hat{q}} \rangle &=  \sum_{y,z} \sum_{i,j} \sum_{(n),(m)}   h_{\ell_1}^{(n)} h_{\ell_1}^{(m)} g_i^y g_j^z \Bigg( \frac{1}{4 \pi} \sum_{\ell_2,\ell_3} (2\ell_2+1)(2\ell_3+1) \begin{pmatrix} \ell_1&\ell_2&\ell_3 \\ 0&0&0 \end{pmatrix}^2 
    \\& \times \Bigg[h_{\ell_2}^{(n)} h_{\ell_2}^{(m)}  \langle C_{\ell_2}^{yz}\rangle
    \langle C_{\ell_3}^{pi(n),qj(m)}\rangle +  h_{\ell_2}^{(n)} h_{\ell_3}^{(m)} \langle C_{\ell_2}^{y, qj(m)}\rangle  \langle C_{\ell_3}^{z, pi(n)}\rangle + h_{\ell_1}^{(n)} h_{\ell_2}^{(m)} \langle \bar{w}^{pi(n)} \rangle \langle b_{\ell_1 \ell_2 \ell_3}^{y,z,qj(m)} \rangle 
    \\&\qquad+ h_{\ell_1}^{(n)} h_{\ell_2}^{(m)} \langle \bar{w}^{qj(m)} \rangle  \langle b_{\ell_1 \ell_2 \ell_3}^{y, z, pi(n)} \rangle + h_{0}^{(n)} h_{\ell_2}^{(m)} \langle \bar{y} \rangle  \langle b_{\ell_1 \ell_2 \ell_3}^{pi(n),z,qj(m)} \rangle + h_{\ell_2}^{(n)} h_{0}^{(m)} \langle \bar{z} \rangle  \langle b_{\ell_1 \ell_2 \ell_3}^{pi(n),y,qj(m)} \rangle \Bigg] 
    \\&+ \frac{1}{2\ell_1+1} \sum_{\ell_2 \ell_3 \ell_4 \ell_5} h_{\ell_2}^{(n)} h_{\ell_3}^{(m)}  \langle \hat{\rho}[y,pi(n),z,qj(m)]^{\ell_2 \ell_4}_{\ell_3 \ell_5}(\ell_1) \rangle \Bigg) \, ,
    \end{aligned}
    }
\end{equation}
where we have used $y$ and $z$ as shorthand for $a^y(\hat{n})$ and $a^z(\hat{n})$, and have also used the fact that $\frac{\langle a_{00} \rangle }{\sqrt{4\pi}}=\langle \bar{a} \rangle$ with $\bar{a}$ denoting the mean of map $a$. Eq.~\eqref{eq.NILC_PS_final} is the main result of this work.

We note that an alternate, simpler analytic form of Eq.~\eqref{eq.NILC_PS_final} is 
\begin{equation}
    \langle C_{\ell_1}^{\hat{p} \hat{q}} \rangle = \frac{1}{2\ell_1+1} \sum_{y,z} \sum_{i,j} \sum_{(n),(m)} \sum_{\ell_2 \ell_3 \ell_4 \ell_5}  h_{\ell_1}^{(n)} h_{\ell_1}^{(m)} g_i^y g_j^z  h_{\ell_2}^{(n)} h_{\ell_3}^{(m)}  \langle \hat{\mathrm{P}}[y,pi(n),z,qj(m)]^{\ell_2 \ell_4}_{\ell_3 \ell_5}(\ell_1) \rangle  \, ,
    \label{eq.NILC_PS_four_point}
\end{equation}
where $\hat{\mathrm{P}}$ is analogous to $\hat{\rho}$ but for the \textit{full} unnormalized trispectrum, as opposed to just the \textit{connected} unnormalized trispectrum. However, computationally, it will be much more efficient to implement Eq.~\eqref{eq.NILC_PS_final} for reasons that will be discussed in Sec.~\ref{sec:computational_implementation}.

As previously mentioned, the derivation of the higher-order terms in Eq.~\eqref{eq.NILC_PS_final} assumes that the weight maps are isotropic in regions that are correlated with the input frequency maps. For many fields of interest, such as the CMB and tSZ effect, this assumption is valid. However, one could imagine a set-up involving only CMB, dust, and instrumental noise, and then using NILC to construct a cleaned CMB map. In such a situation, the weight maps would be correlated with the anisotropic dust field, as the NILC acts to suppress the dust contamination in the final CMB map. The higher-order terms in Eq.~\eqref{eq.NILC_PS_final} would then not be exact, although likely still a very good approximation.


\subsection{Derivation of the Cross-Spectrum of a NILC Map with an Arbitrary Map}

In this subsection, we derive an expression for the contributions to the cross-power spectrum of a NILC map with some other arbitrary map (which is not a NILC map). This is not necessary for, e.g., likelihood analysis of a NILC map power spectrum, but may be useful for other purposes. We denote the harmonic coefficients of the arbitrary map as $T'_{\ell_1 m_1}$ and the harmonic coefficients of the NILC map as $\hat{T}_{\ell_1 m_1}$. Then using Eq.~\eqref{eq.NILC_map}, we seek to compute the following:
\begin{align}
    \langle C_{\ell_1}^{\hat{T}T'} \rangle 
    &= \frac{1}{2\ell_1+1} \sum_{m_1=-\ell_1}^{\ell_1} \langle \hat{T}_{\ell_1 m_1} {T'}^{*}_{\ell_1 m_1} \rangle \nonumber \\
    &= \frac{1}{2\ell_1+1} \sum_{m_1=-\ell_1}^{\ell_1} \sum_{i} \sum_{(n)} \sum_{\ell_2 m_2} \sum_{\ell_4 m_4} (-1)^{m_1} \mathcal{G}^{\ell_1 \ell_2 \ell_4}_{-m_1 m_2 m_4} h^{(n)}_{\ell_1} h^{(n)}_{\ell_2} \langle W^{i(n)}_{\ell_4 m_4} T^{i}_{\ell_2 m_2} {T'}^*_{\ell_1 m_1} \rangle \, ,
    \label{eq.NILC_cross_comp_prelim}
\end{align}
where $\hat{T}$ denotes the NILC map (the same as $T^{\mathrm{NILC}}$ in Eq.~\eqref{eq.NILC_map}).

A subset of this problem was solved in Appendix B of Ref.~\cite{remastered}, which found that the cross-spectrum of a masked map and the (unmasked) map itself is given by
\begin{align}
    \label{eq.atildea}
    \langle C_{\ell_1}^{\tilde{a} a} \rangle 
    &= \frac{1}{2\ell_1+1}\sum_{m_1=-\ell_1}^{\ell_1}\sum_{\ell_2 m_2} \sum_{\ell_4 m_4} (-1)^{m_1} \mathcal{G}^{\ell_1 \ell_2 \ell_4}_{-m_1 m_2 m_4} \langle w_{\ell_4 m_4} a_{\ell_2 m_2}  a^{*}_{\ell_1 m_1} \rangle  \nonumber \\
    &= \frac{1}{4\pi} \sum_{\ell_2,\ell_3} (2\ell_2+1)(2\ell_3+1) \begin{pmatrix} \ell_1&\ell_2&\ell_3 \\ 0&0&0 \end{pmatrix}^2 \langle b^{aaw}_{\ell_1 \ell_2 \ell_3} \rangle  + \frac{\langle w_{00} \rangle}{\sqrt{4\pi}} \langle C_{\ell_1}^{aa} \rangle + \frac{\langle a_{00} \rangle}{\sqrt{4\pi}} \langle C_{\ell_1}^{aw} \rangle \, , 
\end{align}
where $a$, $w$, and $\tilde{a}$ denote a map, mask, and masked map, respectively. The first difference here is that we have two different maps $T$ and $T'$. We thus generalize the result in Eq.~\eqref{eq.atildea} as follows:
\begin{equation}
    \langle C_{\ell_1}^{\tilde{a}a'} \rangle = \frac{1}{4\pi} \sum_{\ell_2,\ell_3} (2\ell_2+1)(2\ell_3+1) \begin{pmatrix} \ell_1&\ell_2&\ell_3 \\ 0&0&0 \end{pmatrix}^2 \langle b^{a'aw}_{\ell_1 \ell_2 \ell_3} \rangle  + \frac{\langle w_{00} \rangle}{\sqrt{4\pi}} \langle C_{\ell_1}^{aa'} \rangle + \frac{\langle a_{00} \rangle}{\sqrt{4\pi}} \langle C_{\ell_1}^{a'w} \rangle \, ,
\end{equation}
where $a'$ is the second map. Applying this result to Eq.~\eqref{eq.NILC_cross_comp_prelim}, and noting that the latter needlet filter's subscript must match that of the original map it filtered (and the $\ell$ subscripts of these maps may have changed due to Kronecker delta functions and index relabeling), we obtain
\begin{align}
    \langle C_{\ell_1}^{\hat{T}T'} \rangle &=  \sum_{i,(n)} h^{(n)}_{\ell_1}  \Bigg( \frac{1}{4\pi} h^{(n)}_{\ell_2} \sum_{\ell_2,\ell_3} (2\ell_2+1)(2\ell_3+1) \begin{pmatrix} \ell_1&\ell_2&\ell_3 \\ 0&0&0 \end{pmatrix}^2 \langle b^{T' T^i W^{i(n)}}_{\ell_1 \ell_2 \ell_3} \rangle  \nonumber 
    \\ &\qquad \qquad \qquad + \frac{\langle w^{i(n)}_{00} \rangle}{\sqrt{4\pi}} h^{(n)}_{\ell_1} \langle C_{\ell_1}^{T^i T'} \rangle + \frac{\langle T^i_{00} \rangle}{\sqrt{4\pi}} h^{(n)}_{0} \langle C_{\ell_1}^{T' W^{i(n)}} \rangle \Bigg) \,.
\end{align}
Finally, noting again that $T^i$ can be represented as a sum over the contributions from all sky components, $T^i = \sum_y g_i^y a^y(\hat{n})$, we thus obtain
\begin{equation}\label{eq.NILC_cross_comp_final}
    \boxed{
    \begin{aligned}
    \langle C_{\ell_1}^{\hat{T}T'} \rangle &=  \sum_{i} \sum_{(n)} h^{(n)}_{\ell_1} g^y_i \Bigg( \bar{w}^{i(n)} h^{(n)}_{\ell_1} \langle C_{\ell_1}^{y T'} \rangle + \bar{y} h^{(n)}_{0} \langle C_{\ell_1}^{T' W^{i(n)}} \rangle 
    \\ &\qquad  +  \frac{1}{4\pi} h^{(n)}_{\ell_2} \sum_{\ell_2,\ell_3} (2\ell_2+1)(2\ell_3+1) \begin{pmatrix} \ell_1&\ell_2&\ell_3 \\ 0&0&0 \end{pmatrix}^2 \langle b^{T' y W^{i(n)}}_{\ell_1 \ell_2 \ell_3} \rangle\, \Bigg) ,
    \end{aligned}}
\end{equation}
where $y$ is used as shorthand for $a^y(\hat{n})$ and we have used the fact that $\frac{\langle a_{00} \rangle }{\sqrt{4\pi}}=\langle \bar{a} \rangle$ with $\bar{a}$ denoting the mean of map $a$.

\section{Computational Implementation}
\label{sec:computational_implementation}
\subsection{Simulated Sky Model}
\label{sec:sky_model}

\begin{figure}[t]
    \centering
    \includegraphics[width=0.8\textwidth]{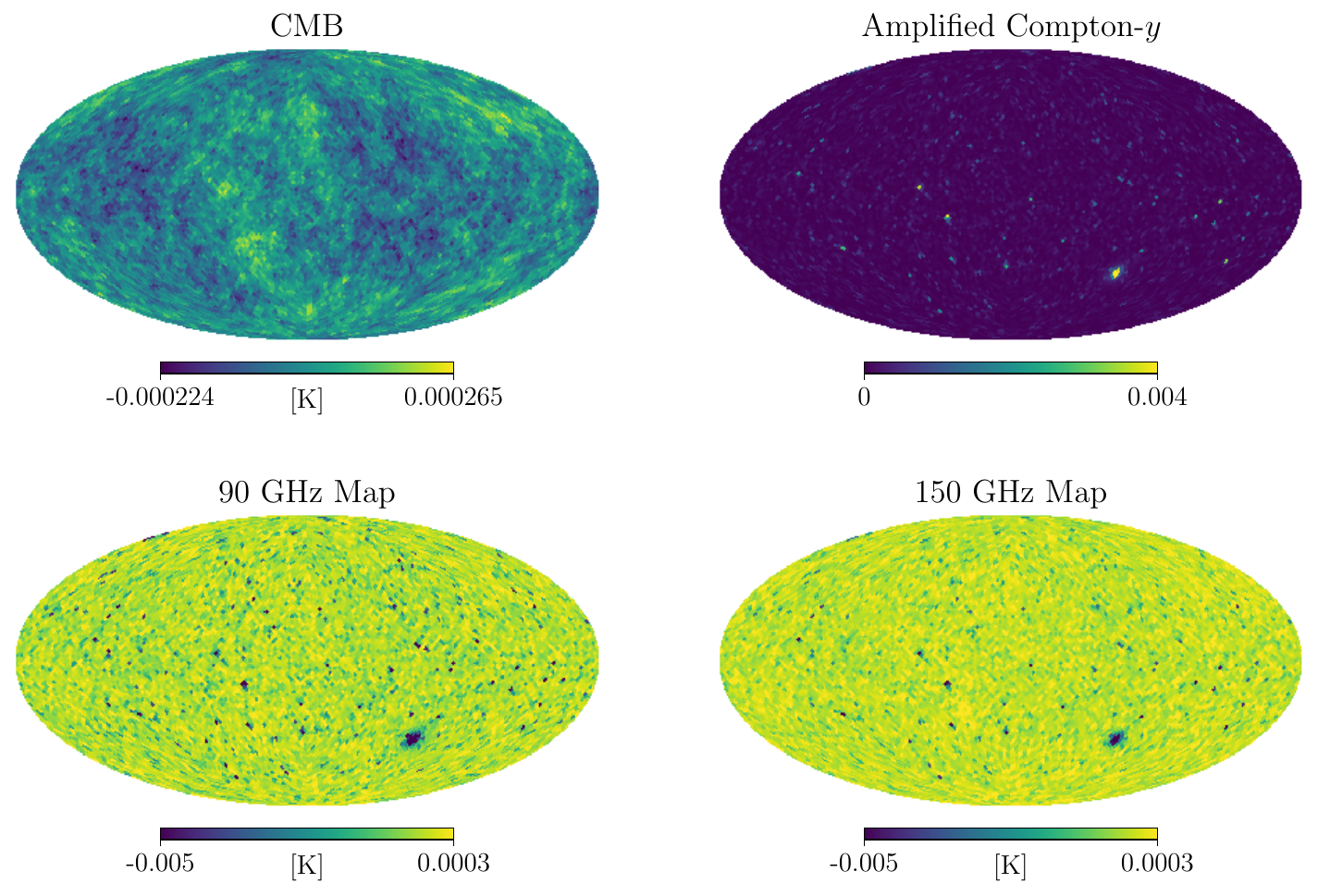}
    \caption{Input CMB and amplified Compton-$y$ map, as well as the total sky map at each frequency. The total sky map consists of CMB, amplified tSZ signal, and instrumental noise. All maps are shown in units of K, except for the Compton-$y$ map, which is dimensionless.}
    \label{fig:input_maps}
\end{figure}

\begin{figure}[t]
    \centering
    \includegraphics[width=0.65\textwidth]{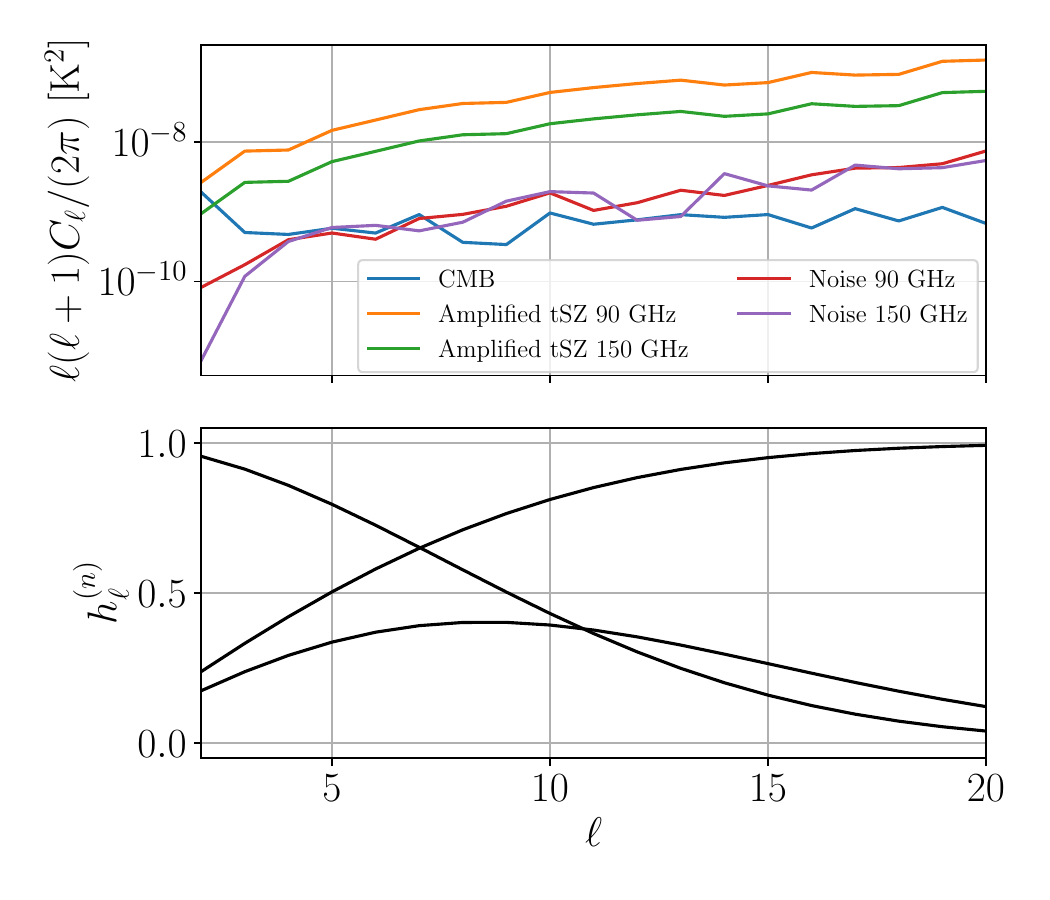}
    \caption{\textbf{Top:} The power spectra, plotted as $\ell(\ell+1)C_\ell/(2\pi)$, of each component at each frequency used in this work. Note that the tSZ field has been amplified by a factor of 1000 at the map level, so as to provide a large non-Gaussian contaminant to the CMB on large angular scales. \textbf{Bottom:} Needlet filters used in this work. These filters are found by taking the difference of two Gaussians, where the Gaussians have FWHM of 1000 and 800 arcmin.}
    \label{fig:spectra_and_filters}
\end{figure}

As a simple demonstrative example, we consider a simulated sky model in which there are two (independent) signals: the thermal Sunyaev-Zel'dovich (tSZ) effect \cite{SZ1969, SZ1970, Birkinshaw_1999} and the CMB, along with noise. We assume that we have only two frequency maps with frequencies of 90 and 150 GHz. Moreover, we amplify the tSZ signal by a factor of 1000 at the map level and refer to it as the ``fake" tSZ signal or ``ftSZ". This is done for demonstration purposes to show how a large non-Gaussian foreground propagates to NILC maps.

Lensed $a_{\ell m}$ for a single CMB realization are obtained from the WebSky Extragalactic CMB Mocks\footnote{\url{https://mocks.cita.utoronto.ca/index.php/WebSky_Extragalactic_CMB_Mocks}} \cite{Websky_2020}. To obtain a realistic, non-Gaussian tSZ map realization, the \verb|halosky|\footnote{\url{https://github.com/marcelo-alvarez/halosky}} package is used. The code works by Poisson sampling from the Tinker et al.~(2008) halo mass function \cite{Tinker:2008} (using the WebSky \cite{Websky_2020} linear matter power spectrum based on the \emph{Planck} 2018 cosmological parameters \cite{Planck:2018cosmoparams} to determine halo abundance) and populating a catalog of halos along the lightcone. We use redshift limits $0 \leq z \leq 5.0$ and mass limits $5 \times 10^{14} M_{\odot} \leq M \leq 10^{16} M_{\odot}$ in this construction, the latter chosen simply for computational efficiency. The pressure profile used is the Battaglia et al.~(2012) AGN feedback profile \cite{Battaglia_2012}. The tSZ map is amplified by a factor of 1000 for our simulations, since we are interested in demonstrating the validity of our analytic results in the presence of large non-Gaussian components, for which the higher-point functions in Eqs.~\eqref{eq.NILC_PS_final} and~\eqref{eq.NILC_cross_comp_final} become significant. For the noise power spectrum, we use the standard model given by Ref.~\cite{Knox1995}: 

\begin{equation*}
    \label{eq.Nell}
    N_\ell = W^2 e^{\ell(\ell+1)\sigma^2} \qquad \text{with} \qquad \sigma = \theta_{\rm FWHM} / \sqrt{8 \ln 2}\,,
\end{equation*}
where $\theta_{\rm FWHM} = 1.4$ arcmin for both the 90 and 150 GHz beams and $W_{90} = W_{150} = 3 \times 10^4 \, \mu {\rm K}_{\rm CMB} \cdot {\rm arcmin}$ in our simulations. We use these large noise power spectra so that the amplitudes of the noise spectra are comparable to those of the ftSZ power spectra. This is important for our simple two-frequency, two-component-plus-noise sky model, where we do not want the NILC maps to completely clean the contaminating foregrounds (thus giving a realistic representation on what would happen with actual data where there are several sky components that cannot all be simultaneously cleaned completely). The simulated CMB and amplified Compton-$y$ maps (with HEALPix resolution parameter $N_{\mathrm{side}}=32$), as well as the total sky map at each frequency (comprising CMB, amplified tSZ, and instrumental noise), are shown in Fig.~\ref{fig:input_maps}. The power spectra of these simulated maps are shown in the top panel of Fig.~\ref{fig:spectra_and_filters}. We limit ourselves to low $\ell_{\rm max}$ and low map resolution for computational efficiency, since this is simply a validation check of our analytic result. The maps are band-limited at $\ell_{\rm max}=20$ to avoid having to compute sums over arbitrarily high $\ell$ in Eq.~\eqref{eq.NILC_PS_final}.


\subsection{Computational Implementation of NILC Power Spectrum Analytic Result}

\begin{figure}[t]
    \centering
    \includegraphics[width=0.9\textwidth]{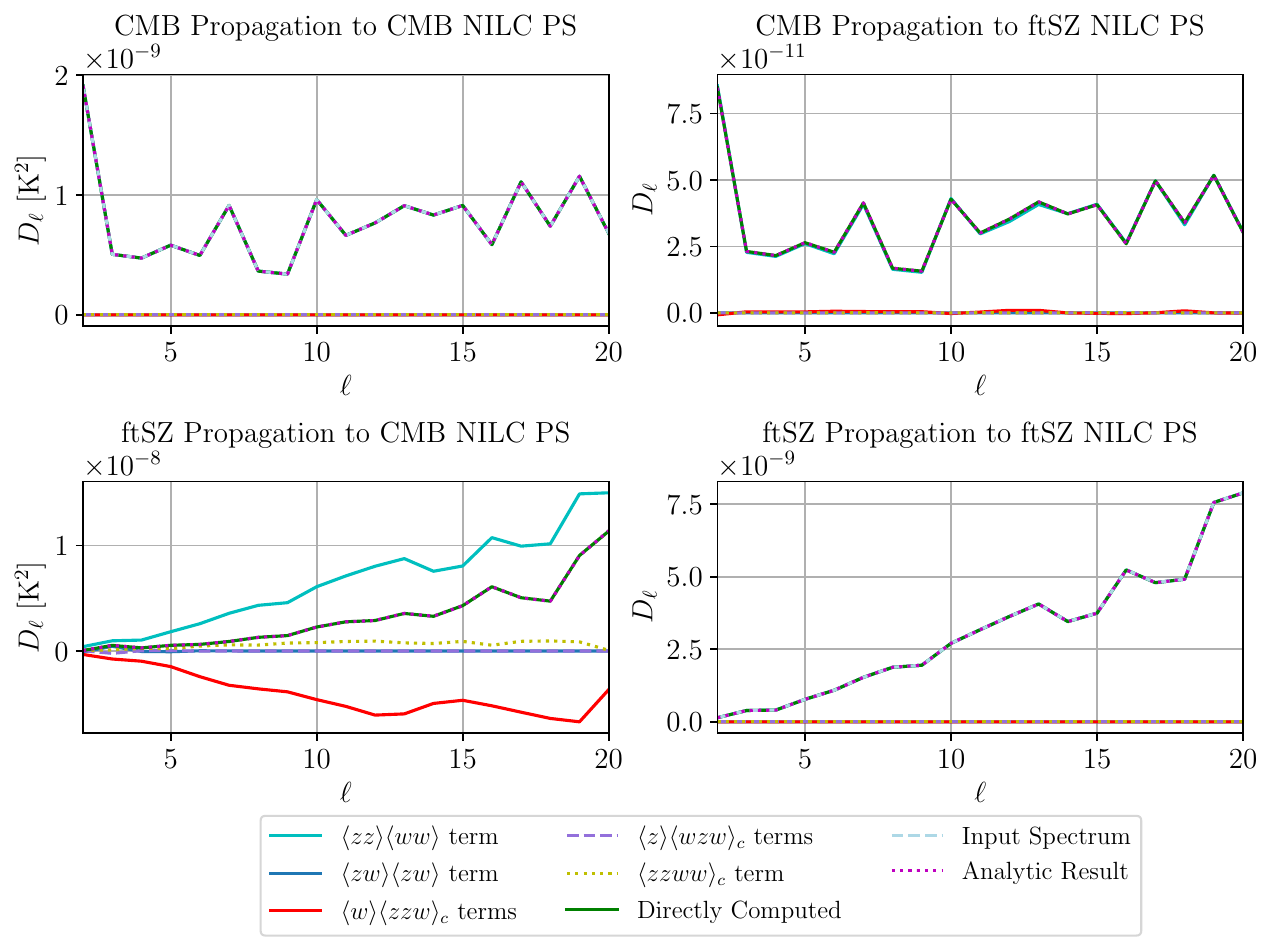}
    \caption{Propagation of components to NILC map auto-spectra. \textbf{Top left}: propagation of the CMB signal of interest to a CMB-preserved NILC map auto-spectrum. \textbf{Top right}: propagation of the CMB contaminant to a ftSZ-preserved NILC map auto-spectrum. \textbf{Bottom left}: propagation of the ftSZ contaminant to a CMB-preserved NILC map auto-spectrum. \textbf{Bottom right}: propagation of the ftSZ signal of interest to a ftSZ-preserved NILC map auto-spectrum. In all plots, the analytic propagation of the component (dotted magenta) is compared to the directly computed (simulation-based) propagation of the component (solid green). As expected, these agree exactly (i.e., the dotted magenta lines lie precisely on top of the solid green lines), thus validating both the analytic result and its computational implementation. Contributions from the various terms in Eq.~\eqref{eq.NILC_PS_final} are also shown individually as labeled. For the CMB propagation to a CMB-preserved NILC and ftSZ propagation to a ftSZ-preserved NILC, we also show the input power spectra (dashed light blue). These match the directly computed and analytic results, verifying that the signal of interest propagates in an unbiased fashion to the final map, as expected in the NILC method. All curves are plotted as $D_\ell = \ell(\ell+1)C_\ell/(2\pi)$, in units of $\mathrm{K}^2$ for the left two panels and in dimensionless Compton-$y$ units for the right two panels. } 
    \label{fig:propagation}
\end{figure}

To validate the result in Eq.~\eqref{eq.NILC_PS_final} and assess the relative importance of the different terms, we compute the propagation of both the signal of interest and the contaminating foreground to the auto-power spectrum of each NILC map (i.e., we consider both a CMB-preserved NILC map and a tSZ-preserved NILC map). Specifically, we rewrite Eq.~\eqref{eq.NILC_PS_final} as 
\begin{equation}
    \langle C_{\ell_1}^{\hat{p}\hat{q}} \rangle = \sum_{y,z} C_\ell^{y \rightarrow \hat{p}, z \rightarrow \hat{q}},
\end{equation} where
\begin{align}
    C_\ell^{y \rightarrow \hat{p}, z \rightarrow \hat{q}} &\equiv  \sum_{i,j} \sum_{(n),(m)}   h_{\ell_1}^{(n)} h_{\ell_1}^{(m)} g_i^y g_j^z \Bigg( \frac{1}{4 \pi} \sum_{\ell_2,\ell_3} (2\ell_2+1)(2\ell_3+1) \begin{pmatrix} \ell_1&\ell_2&\ell_3 \\ 0&0&0 \end{pmatrix}^2 \nonumber
    \\& \times \Bigg[h_{\ell_2}^{(n)} h_{\ell_2}^{(m)}  \langle C_{\ell_2}^{yz}\rangle
    \langle C_{\ell_3}^{pi(n),qj(m)}\rangle +  h_{\ell_2}^{(n)} h_{\ell_3}^{(m)} \langle C_{\ell_2}^{y, qj(m)}\rangle  \langle C_{\ell_3}^{z, pi(n)}\rangle + h_{\ell_1}^{(n)} h_{\ell_2}^{(m)} \langle \bar{w}^{pi(n)} \rangle \langle b_{\ell_1 \ell_2 \ell_3}^{y,z,qj(m)} \rangle \nonumber
    \\\nonumber &\qquad+ h_{\ell_1}^{(n)} h_{\ell_2}^{(m)} \langle \bar{w}^{qj(m)} \rangle  \langle b_{\ell_1 \ell_2 \ell_3}^{y, z, pi(n)} \rangle + h_{0}^{(n)} h_{\ell_2}^{(m)} \langle \bar{y} \rangle  \langle b_{\ell_1 \ell_2 \ell_3}^{pi(n),z,qj(m)} \rangle + h_{\ell_2}^{(n)} h_{0}^{(m)} \langle \bar{z} \rangle  \langle b_{\ell_1 \ell_2 \ell_3}^{pi(n),y,qj(m)} \rangle \Bigg] 
    \\&+ \frac{1}{2\ell_1+1} \sum_{\ell_2 \ell_3 \ell_4 \ell_5} h_{\ell_2}^{(n)} h_{\ell_3}^{(m)}  \langle \hat{\rho}[y,pi(n),z,qj(m)]^{\ell_2 \ell_4}_{\ell_3 \ell_5}(\ell_1) \rangle \Bigg) \, .
    \label{eq.Clypzq_def}
\end{align}
Intuitively, $C_\ell^{y \rightarrow \hat{p}, z \rightarrow \hat{q}}$ is the (cross-)power spectrum of the propagation of component $y$ to a $p$-preserved NILC map and the propagation of component $z$ to a $q$-preserved NILC map. 

There are two ways to compute $C_\ell^{y \rightarrow \hat{p}, z \rightarrow \hat{q}}$. The first is via computation of all the $n$-point functions in Eq.~\eqref{eq.Clypzq_def}. The second is via a simulation-based approach, as follows: Using the full 90 and 150 GHz frequency maps, determine the NILC weight maps for the preserved component $p$. Next, consider only the contributions of component $y$ to the 90 and 150 GHz maps. Using these as the new frequency maps, along with the previously determined weight maps, build the resulting NILC map using the steps in Sec.~\ref{sec:needlet_ILC}. This gives the contribution of component $y$ to a $p$-preserved NILC map. Repeat the procedure for component $z$ and preserved component $q$. Taking the (cross-)power spectrum of the resulting two NILC maps gives $C_\ell^{y \rightarrow \hat{p}, z \rightarrow \hat{q}}$. Below, we compare the results from the analytic $n$-point calculation approach and the simulation-based approach as a demonstration of the validity of Eq.~\eqref{eq.NILC_PS_final}.
 
For our set-up, we use three needlet filter scales. The filters are obtained by taking the differences of successive Gaussians (as in Ref.~\cite{Planck2015ymap}), where our Gaussians have FWHM of 1000 and 800 arcmin, chosen to have differing scale dependence on the large scales for which we perform this analysis. These filters are shown in the bottom panel of Fig.~\ref{fig:spectra_and_filters}. We consider only three filters here for computational efficiency. With more filters, the NILC algorithm is able to capture more scale-dependent and spatially varying information and is thus expected to clean non-Gaussian foregrounds even more effectively. Next, the simulated sky maps at 90 and 150 GHz are run through a pipeline that generates NILC weight maps, \verb|pyilc|\footnote{\url{https://github.com/jcolinhill/pyilc}}  \cite{McCarthy:2023hpa, McCarthy:2023cwg}.

For computation of the bispectra and trispectra we adapt code from \verb|PolyBin|\footnote{\url{https://github.com/oliverphilcox/PolyBin}} \cite{PhilcoxNpoint}, and for calculation of the various terms in Eq.~\eqref{eq.NILC_PS_final} we adapt code from \verb|reMASTERed|\footnote{\url{https://github.com/kmsurrao/reMASTERed}} \cite{remastered}. In most cases of physical interest, the first term in Eq.~\eqref{eq.NILC_PS_final} is dominant. Since calculation of the bispectrum and trispectrum terms is the computational bottleneck, it would be possible to bin in $\ell$ for calculation of these terms in cases where their effects are small. For this reason, we prefer to implement Eq.~\eqref{eq.NILC_PS_final} over Eq.~\eqref{eq.NILC_PS_four_point}, as computing Eq.~\eqref{eq.NILC_PS_four_point} with our desired precision would bar the use of $\ell$-space binning for the full trispectrum calculation, significantly slowing the runtime. However, in the demonstrations here, we do not use any $\ell$-space binning and instead perform the computations only at very low $\ell$, as our goal is primarily to validate our analytic results.

For our demonstrations, we consider the specific case of $y=z$ and $\hat{p}=\hat{q}$, or in other words, how the auto-power spectrum of some component propagates into the auto-power spectrum of some NILC map (note that the propagating component does not have to be the same as the component preserved in the NILC map). In particular, Fig.~\ref{fig:propagation} shows the CMB and ftSZ propagation into CMB and ftSZ NILC map auto-power spectra. As expected, the ``directly computed" simulation-based results (solid green curves) and the analytic results from Eq.~\eqref{eq.NILC_PS_final} (dotted magenta curves) agree, for both the signal of interest and the contaminant. Moreover, as expected, the signal of interest propagates in an unbiased fashion to the resulting NILC map, matching the input power spectrum of that signal (dashed light blue curve in the top left and bottom right panels of Fig.~\ref{fig:propagation}). In those panels, the solid green, dotted magenta, and dashed light blue curves all lie directly on top of each other, as expected (the directly computed [solid green] and analytic [dotted magenta] results should always agree if our analytic results are correct, and the input spectrum [dashed light blue] should match both of those if we consider how a component propagates to its own NILC map, assuming negligible ILC bias). Contributions of various terms (power spectrum, bispectrum, and trispectrum) in Eq.~\eqref{eq.NILC_PS_final} are also plotted, as described in the figure legend, to assess the relative contributions of the terms. Of particular note is the bottom left panel of Fig.~\ref{fig:propagation}, where the red curve involving a bispectrum factor is of comparable magnitude to the cyan curve involving power spectrum factors. This is the case because the non-Gaussian ftSZ field has a non-zero bispectrum. When building a CMB NILC map, the weights aim to downweight the ftSZ contaminant and will thus be correlated with that contaminant. Hence, the bispectrum term involving two factors of the ftSZ field and one factor of the weight map that acts to suppress that field is also non-zero. 

We also note the units of the different panels of the figure. Specifically, the power spectrum of a CMB NILC map in this set-up has units of $\mathrm{K}^2$, and the power spectrum of a NILC Compton-$y$ map is dimensionless. Thus, the contribution to the power spectrum from a component that propagates into a CMB NILC map power spectrum must also be in $\mathrm{K}^2$, while the contribution to the power spectrum from a component that propagates into an ftSZ NILC map power spectrum must be dimensionless. It may be somewhat surprising that the ftSZ propagation into a CMB NILC map has units of $\mathrm{K}^2$, given that we are computing the propagation of the Compton-$y$ field, not the propagation of the ftSZ field at a given frequency (or pair of frequencies). However, looking at Eq.~\eqref{eq.Clypzq_def}, we note that the propagation is actually being summed over all possible frequency pairs to get the total contribution. Thus, the contribution from the ftSZ field comes from all frequencies instead of singling out any individual frequency or pair of frequencies. The code used for these demonstrations is publicly available in \verb|NILC-PS-Model|. We discuss how non-Gaussian information is captured by the NILC process in the next section.


\section{Discussion}
\label{sec.discussion}

In this paper, we have derived an analytic expression for the auto- and cross-spectra of NILC maps in Eq.~\eqref{eq.NILC_PS_final} and an expression for the cross-spectrum of a NILC map with an arbitrary second map in Eq.~\eqref{eq.NILC_cross_comp_final}. Eq.~\eqref{eq.NILC_PS_final} elucidates how NILC power spectra capture non-Gaussian information. The equation involves correlators of the maps and weight maps, including connected three- and four-point functions. Since the NILC weight maps are constructed to suppress contaminants, the weight maps themselves are functions of the contaminant fields. Thus, the higher-point functions of the maps and weight maps contain the non-Gaussian information about the contaminant foregrounds. Fig.~\ref{fig:propagation} demonstrates the relative importance of the various terms. In particular, we note the importance of the connected bispectrum involving two factors of the map and one factor of the weight map. This term appears to play an important role for the propagation of non-Gaussian components, as seen in the bottom left panel of that figure. Comparing with the top right panel, we see that the higher-point functions are significantly less important for propagation of Gaussian random fields, as one would expect.

Notably, for the signal of interest to propagate in an unbiased fashion to a NILC map, only the first term in Eq.~\eqref{eq.NILC_PS_final} can be non-zero when either $y$ or $z$ represents the component of interest. As long as the local real-space domains on which the needlet ILC is performed are large enough, the weight maps should not be correlated with the signal of interest, forcing the other terms to zero, as required. However, if the size of these domains is too small, there may be correlations of the signal field with the weight maps, biasing the way the signal propagates into the final NILC map. This is equivalent to the usual ILC bias that results from computing the covariances for ILC weights using an insufficient number of modes \cite{Delabrouille2009}. In our demonstrations, \verb|pyilc| allows the user to specify an ILC bias tolerance, which we set to 1\%. This bias tolerance determines how large the real-space domains used for computing covariance matrices in the NILC algorithm need to be in order to keep the bias below that threshold. Fig.~\ref{fig:propagation} demonstrates the lack of ILC bias in our demonstration, as each component propagates in an unbiased fashion to the NILC map which preserves that component, matching the input power spectrum of the signal (see top left and bottom right panels). As required, only the first term in Eq.~\eqref{eq.NILC_PS_final} is non-zero in that case.

Because the weight maps are functions of the fields, our analytic results demonstrate that the parameter dependence of NILC power spectra is nonlinear, when the input fields are parameterized via an overall amplitude parameter for each component power spectrum. Parameterizing a component at the power spectrum level alone is no longer sufficient to describe that component's contribution to a NILC map power spectrum. Instead, one needs a model that (at least) describes the two-, three-, and four-point functions (and thus, ideally, a field-level model). In some cases, we find that when the amplitude of a contaminant increases, its contribution to the power spectrum of a cleaned NILC map actually decreases since the weight maps then prioritize suppressing that component over the others. Because of this complicated parameter dependence, it is unfortunately not obvious how to use Eq.~\eqref{eq.NILC_PS_final} for performing cosmological parameter inference analytically. In paper II of this work~\cite{PaperII}, we numerically determine the parameter dependence of NILC power spectra by computing their derivatives with respect to the amplitude of each input component map and performing symbolic regression. We also investigate likelihood-free inference, where the parameter dependence is learned implicitly via normalizing flows. In particular, since the Gaussian likelihood would be inaccurate in cases of interest here, likelihood-free inference is technically necessary. We use neural posterior estimation for this purpose \cite{SNPE}. This analysis will be described in our companion work~\cite{PaperII}.

Because full fields need to be simulated to obtain weight maps that contain spatial information, and because, currently, it is significantly slower to compute component propagation to a NILC map via our $n$-point function result than it is to compute it directly using simulations, these analytic results are intractable to use directly in cosmological parameter inference pipelines at this time. Nevertheless, our analytic results clearly demonstrate the ability of NILC to capture non-Gaussian information, and furthermore provide a significant step forward in understanding the sky component contributions contained in NILC maps.

\section{Acknowledgments}
We thank Oliver Philcox for help with bispectrum and trispectrum estimation; Marcelo Alvarez for help with generating tSZ simulations with \verb|halosky|; and William Coulton and Fiona McCarthy for useful discussions. This material is based upon work supported by the National Science Foundation Graduate Research Fellowship Program under Grant No.~DGE 2036197 (KMS). JCH acknowledges support from NSF grant AST-2108536, NASA grants 80NSSC23K0463 and 80NSSC22K0721, the Sloan Foundation, and the Simons Foundation. Several software tools were used in the development and presentation of results shown in this paper, including \verb|HEALPix/healpy| \cite{Healpix, Healpy}, \verb|numpy| \cite{numpy}, \verb|scipy| \cite{scipy}, \verb|matplotlib| \cite{matplotlib}, \verb|astropy| \cite{astropy1, astropy2, astropy3}, and \verb|pywigxjpf| \cite{pywigxjpf}. The authors acknowledge the Texas Advanced Computing Center (TACC) at The University of Texas at Austin for providing HPC resources that have contributed to the research results reported within this paper. The authors also acknowledge the use of resources of the National Energy Research Scientific Computing Center (NERSC), a U.S.~Department of Energy Office of Science User Facility located at Lawrence Berkeley National Laboratory. Finally, the authors acknowledge the use of computing resources from Columbia University's Shared Research Computing Facility project, which is supported by NIH Research Facility Improvement Grant 1G20RR030893-01, and associated funds from the New York State Empire State Development, Division of Science Technology and Innovation (NYSTAR) Contract C090171, both awarded April 15, 2010. 

\begin{appendices}

\section{Generalized Result for the Cross-Spectrum of Two Masked Maps}
\label{app:reMASTERed_generalization}

In this appendix we derive the expansion of the cross-power spectrum of two masked (or weighted) maps in terms of $n$-point functions, the results of which are given in \S \ref{sec:contributions}. We refer the reader to Appendix A of \cite{remastered} for the full derivation of the auto-spectrum of a masked map. We closely follow that derivation but use $a$ and $a'$ to denote two maps, $w$ and $w'$ to denote their respective masks, and $\tilde{a}$ and $\tilde{a}'$ to denote the respective masked maps. Then the ensemble-averaged masked map cross-spectrum is given by
\begin{align}
    \langle C^{\tilde{a},\tilde{a}'}_{\ell_1} \rangle &\equiv \frac{1}{2\ell_1+1}\sum_{m_1=-\ell_1}^{\ell_1} \langle \tilde{a}_{\ell_1m_1}  (\tilde{a}'_{\ell_1m_1})^* \rangle \nonumber
    \\&= \frac{1}{2\ell_1+1}\sum_{m_1=-\ell_1}^{\ell_1}\sum_{\ell_2 m_2}\sum_{\ell_3 m_3} \sum_{\ell_4 m_4} \sum_{\ell_5 m_5} \langle a_{\ell_2 m_2}(a'_{\ell_3 m_3})^* w_{\ell_4 m_4} (w'_{\ell_5 m_5})^* \rangle \mathcal{G}^{\ell_1 \ell_2 \ell_4}_{-m_1 m_2 m_4} \mathcal{G}^{\ell_1 \ell_3 \ell_5}_{m_1 -m_3 -m_5}  \nonumber
    \\&= \frac{1}{2\ell_1+1}\sum_{m_1=-\ell_1}^{\ell_1}\sum_{\ell_2 m_2}\sum_{\ell_3 m_3} \sum_{\ell_4 m_4} \sum_{\ell_5 m_5} (-1)^{m_1} \langle a_{\ell_2 m_2}a'_{\ell_3 m_3} w_{\ell_4 m_4} w'_{\ell_5 m_5} \rangle \mathcal{G}^{\ell_1 \ell_2 \ell_4}_{-m_1 m_2 m_4} \mathcal{G}^{\ell_1 \ell_3 \ell_5}_{m_1 m_3 m_5} \nonumber
    \\&= \frac{1}{2\ell_1+1} \sum_{m_1=-\ell_1}^{\ell_1}\sum_{\ell_2 m_2}\sum_{\ell_3 m_3} \sum_{\ell_4 m_4} \sum_{\ell_5 m_5} (-1)^{m_1} \mathcal{G}^{\ell_1 \ell_2 \ell_4}_{-m_1 m_2 m_4} \mathcal{G}^{\ell_1 \ell_3 \ell_5}_{m_1 m_3 m_5} \nonumber
    [ \langle a_{\ell_2 m_2}a'_{\ell_3m_3}\rangle  \langle w_{\ell_4 m_4}w'_{\ell_5m_5}\rangle \notag\\ &\qquad + \langle a_{\ell_2 m_2}w_{\ell_4m_4} \rangle \langle a'_{\ell_3 m_3}w'_{\ell_5m_5} \rangle + \langle a_{\ell_2 m_2}w'_{\ell_5m_5} \rangle \langle a'_{\ell_3 m_3}w_{\ell_4m_4} \rangle + \langle w_{\ell_4 m_4} \rangle \langle a_{\ell_2 m_2} a'_{\ell_3 m_3} w'_{\ell_5 m_5} \rangle_c \nonumber \notag\\ &\qquad + \langle w'_{\ell_5 m_5} \rangle \langle a_{\ell_2 m_2} a'_{\ell_3 m_3} w_{\ell_4 m_4} \rangle_c + \langle a_{\ell_2 m_2} \rangle \langle a'_{\ell_3 m_3} w_{\ell_4 m_4} w'_{\ell_5 m_5} \rangle_c + \langle a'_{\ell_3 m_3} \rangle \langle a_{\ell_2 m_2} w_{\ell_4 m_4} w'_{\ell_5 m_5}\rangle_c \nonumber \notag\\ &\qquad  + \langle a_{\ell_2 m_2} a'_{\ell_3 m_3} w_{\ell_4 m_4} w'_{\ell_5 m_5} \rangle_c],
    \label{eq.appendix_full_expansion}
\end{align}
where we have written out the Wick contractions of the four-point function in the final line. Using definitions of the power spectrum, bispectrum, and trispectrum,  
\begin{align}
    \langle C^{\tilde{a},\tilde{a}'}_{\ell_1} \rangle &= \frac{1}{2\ell_1+1} \sum_{m_1=-\ell_1}^{\ell_1}\sum_{\ell_2 m_2}\sum_{\ell_3 m_3} \sum_{\ell_4 m_4} \sum_{\ell_5 m_5} (-1)^{m_1} \mathcal{G}^{\ell_1 \ell_2 \ell_4}_{-m_1 m_2 m_4} \mathcal{G}^{\ell_1 \ell_3 \ell_5}_{m_1 m_3 m_5} \bigg[ 
    \nonumber \notag\\ &\qquad \langle C_{\ell_2}^{aa'}\rangle  \langle C_{\ell_4}^{ww'}\rangle (-1)^{m_2+m_4} \delta^{\rm K}_{\ell_2, \ell_3}\delta^{\rm K}_{m_2,-m_3}\delta^{\rm K}_{\ell_4, \ell_5}\delta^{\rm K}_{m_4,-m_5} + \langle C_{\ell_2}^{aw} \rangle \langle C_{\ell_3}^{a'w'} \rangle (-1)^{m_2+m_3} \delta^{\rm K}_{\ell_2, \ell_4}\delta^{\rm K}_{m_2,-m_4}\delta^{\rm K}_{\ell_3, \ell_5}\delta^{\rm K}_{m_3,-m_5} 
     \nonumber \notag\\ &\qquad + \langle C_{\ell_2}^{aw'} \rangle \langle C_{\ell_3}^{a'w} \rangle (-1)^{m_2+m_3} \delta^{\rm K}_{\ell_2, \ell_5}\delta^{\rm K}_{m_2,-m_5}\delta^{\rm K}_{\ell_3, \ell_4}\delta^{\rm K}_{m_3,-m_4} 
    \nonumber \notag\\ &\qquad + \langle w_{\ell_4 m_4} \rangle \mathcal{G}^{\ell_2 \ell_3 \ell_5}_{m_2 m_3 m_5} b_{\ell_2 \ell_3 \ell_5}^{aa'w'} +  \langle w'_{\ell_5 m_5} \rangle \mathcal{G}^{\ell_2 \ell_3 \ell_4}_{m_2 m_3 m_4} b_{\ell_2 \ell_3 \ell_4}^{aa'w} + \langle a_{\ell_2 m_2} \rangle G^{\ell_3 \ell_4 \ell_5}_{m_3 m_4 m_5} b^{a'ww'}_{\ell_3 \ell_4 \ell_5} + \langle a'_{\ell_3 m_3} \rangle G^{\ell_2 \ell_4 \ell_5}_{m_2 m_4 m_5} b^{aww'}_{\ell_2 \ell_4 \ell_5}  \nonumber \notag\\ &\qquad + \sum_{L=0}^{\infty} \sum_{M=-L}^{L} (-1)^M \mathcal{G}^{\ell_2 \ell_3 L}_{m_2 m_3 -M} \mathcal{G}^{\ell_4 \ell_5 L}_{m_4 m_5 M} t[aa'ww']^{\ell_2 \ell_3}_{\ell_4 \ell_5}(L) + \text{ 23 perms.} \bigg],
\end{align}
where we have assumed isotropy of the mask in regions correlated with the fields. Applying the Kronecker deltas gives

\begin{align}
    \label{eq.eight_terms}
    \langle \tilde{C}_{\ell_1} \rangle &= \frac{1}{2\ell_1+1} \sum_{m_1=-\ell_1}^{\ell_1}\sum_{\ell_2 m_2}  \sum_{\ell_4 m_4} (-1)^{m_1+m_2+m_4} \mathcal{G}^{\ell_1 \ell_2 \ell_4}_{-m_1 m_2 m_4} \mathcal{G}^{\ell_1 \ell_2 \ell_4}_{m_1 -m_2 -m_4} \langle C_{\ell_2}^{aa'}\rangle  \langle C_{\ell_4}^{ww'}\rangle  
    \\\nonumber \notag &\;\; + \frac{1}{2\ell_1+1} \sum_{m_1=-\ell_1}^{\ell_1}\sum_{\ell_2 m_2} \sum_{\ell_3 m_3} (-1)^{m_1+m_2+m_3} \mathcal{G}^{\ell_1 \ell_2 \ell_2}_{-m_1 m_2 -m_2} \mathcal{G}^{\ell_1 \ell_3 \ell_3}_{m_1 m_3 -m_3}   \langle C_{\ell_2}^{aw}\rangle  \langle C_{\ell_3}^{a'w'}\rangle     
    \nonumber \notag\\ &\;\; + \frac{1}{2\ell_1 + 1} \sum_{m_1=-\ell_1}^{\ell_1}\sum_{\ell_2 m_2} \sum_{\ell_3 m_3} (-1)^{m_1+m_2+m_3} \mathcal{G}^{\ell_1 \ell_2 \ell_3}_{-m_1 m_2 -m_3} \mathcal{G}^{\ell_1 \ell_3 \ell_2}_{m_1 m_3 -m_2} \langle C_{\ell_2}^{aw'}\rangle  \langle C_{\ell_3}^{a'w}\rangle 
    \nonumber \notag\\ &\;\; + \frac{1}{2\ell_1+1} \sum_{m_1=-\ell_1}^{\ell_1}\sum_{\ell_2 m_2}\sum_{\ell_3 m_3} \sum_{\ell_4 m_4} \sum_{\ell_5 m_5} (-1)^{m_1} \mathcal{G}^{\ell_1 \ell_2 \ell_4}_{-m_1 m_2 m_4} \mathcal{G}^{\ell_1 \ell_3 \ell_5}_{m_1 m_3 m_5} \mathcal{G}^{\ell_2 \ell_3 \ell_5}_{m_2 m_3 m_5} \langle w_{\ell_4 m_4} \rangle  \langle b_{\ell_2 \ell_3 \ell_5}^{aa'w'} \rangle 
    \nonumber \notag\\ &\;\; + \frac{1}{2\ell_1+1} \sum_{m_1=-\ell_1}^{\ell_1}\sum_{\ell_2 m_2}\sum_{\ell_3 m_3} \sum_{\ell_4 m_4} \sum_{\ell_5 m_5} (-1)^{m_1} \mathcal{G}^{\ell_1 \ell_2 \ell_4}_{-m_1 m_2 m_4} \mathcal{G}^{\ell_1 \ell_3 \ell_5}_{m_1 m_3 m_5} \mathcal{G}^{\ell_2 \ell_3 \ell_4}_{m_2 m_3 m_4} \langle w'_{\ell_5 m_5} \rangle  \langle b_{\ell_2 \ell_3 \ell_4}^{aa'w} \rangle 
    \nonumber \notag\\ &\;\; + \frac{1}{2\ell_1+1} \sum_{m_1=-\ell_1}^{\ell_1}\sum_{\ell_2 m_2}\sum_{\ell_3 m_3} \sum_{\ell_4 m_4} \sum_{\ell_5 m_5} (-1)^{m_1} \mathcal{G}^{\ell_1 \ell_2 \ell_4}_{-m_1 m_2 m_4} \mathcal{G}^{\ell_1 \ell_3 \ell_5}_{m_1 m_3 m_5} \mathcal{G}^{\ell_3 \ell_4 \ell_5}_{m_3 m_4 m_5} \langle a_{\ell_2 m_2} \rangle  \langle b_{\ell_3 \ell_4 \ell_5}^{a'ww'} \rangle 
    \nonumber \notag\\ &\;\; + \frac{1}{2\ell_1+1} \sum_{m_1=-\ell_1}^{\ell_1}\sum_{\ell_2 m_2}\sum_{\ell_3 m_3} \sum_{\ell_4 m_4} \sum_{\ell_5 m_5} (-1)^{m_1} \mathcal{G}^{\ell_1 \ell_2 \ell_4}_{-m_1 m_2 m_4} \mathcal{G}^{\ell_1 \ell_3 \ell_5}_{m_1 m_3 m_5} \mathcal{G}^{\ell_2 \ell_4 \ell_5}_{m_2 m_4 m_5} \langle a'_{\ell_3 m_3} \rangle  \langle b_{\ell_2 \ell_4 \ell_5}^{aww'} \rangle 
    \nonumber \notag\\ &\;\; + \frac{1}{2\ell_1+1} \sum_{m_1=-\ell_1}^{\ell_1}\sum_{\ell_2 m_2}\sum_{\ell_3 m_3} \sum_{\ell_4 m_4} \sum_{\ell_5 m_5} \sum_{LM} (-1)^{m_1+M} \mathcal{G}^{\ell_1 \ell_2 \ell_4}_{-m_1 m_2 m_4} \mathcal{G}^{\ell_1 \ell_3 \ell_5}_{m_1 m_3 m_5} \mathcal{G}^{\ell_2 \ell_3 L}_{m_2 m_3 -M} \mathcal{G}^{\ell_4 \ell_5 L}_{m_4 m_5 M} t[aa'ww']^{\ell_2 \ell_3}_{\ell_4 \ell_5}(L) 
    \nonumber \notag\\ &\hspace{390 pt} + \text{ 23 perms.}\nonumber,
    \end{align}
where the permutations are taken over just the final two Gaunt factors and reduced trispectrum in the last line. We refer to each of the above terms schematically as $\langle aa \rangle \langle ww \rangle$, $\langle aw \rangle \langle aw \rangle$, $\langle w \rangle \langle aaw \rangle_c$, $\langle a \rangle \langle waw \rangle_c$, and $\langle aaww \rangle_c$. 

 Using properties of the Wigner symbols, we simplify each of the terms in Eq.~\eqref{eq.eight_terms} separately.\footnote{\url{https://functions.wolfram.com/HypergeometricFunctions/ThreeJSymbol/}} First consider the $\langle aa \rangle \langle ww \rangle$ term:
\begin{align}
    &\frac{1}{2\ell_1+1} \sum_{m_1=-\ell_1}^{\ell_1}\sum_{\ell_2 m_2} \sum_{\ell_4 m_4} (-1)^{m_1+m_2+m_4} \mathcal{G}^{\ell_1 \ell_2 \ell_4}_{-m_1 m_2 m_4} \mathcal{G}^{\ell_1 \ell_2 \ell_4}_{m_1 -m_2 -m_4} \langle C_{\ell_2}^{aa'}\rangle  \langle C_{\ell_4}^{ww'}\rangle  \nonumber
    \\&= \frac{1}{4\pi} \sum_{\ell_2,\ell_3} (2\ell_2+1)(2\ell_3+1) \begin{pmatrix} \ell_1&\ell_2&\ell_3 \\ 0&0&0 \end{pmatrix}^2 \langle C_{\ell_2}^{aa'}\rangle  \langle C_{\ell_3}^{ww'}\rangle \label{eq.term1_lastline} \,.
\end{align}
This is the analog of the usual term computed in the MASTER formalism~\cite{Hivon2002}. The first $\langle aw \rangle \langle aw \rangle$ term from Eq.~\eqref{eq.eight_terms} becomes
\begin{align}
    &\frac{1}{2\ell_1+1} \sum_{m_1=-\ell_1}^{\ell_1}\sum_{\ell_2 m_2} \sum_{\ell_3 m_3} (-1)^{m_1+m_2+m_3} \mathcal{G}^{\ell_1 \ell_2 \ell_2}_{-m_1 m_2 -m_2} \mathcal{G}^{\ell_1 \ell_3 \ell_3}_{m_1 m_3 -m_3}  \langle C_{\ell_2}^{aw}\rangle  \langle C_{\ell_3}^{a'w'}\rangle  \nonumber
    \\&= \frac{1}{2\ell_1+1} \sum_{\ell_2 m_2} \sum_{\ell_3 m_3} (-1)^{m_2+m_3} \mathcal{G}^{\ell_1 \ell_2 \ell_2}_{0 m_2 -m_2} \mathcal{G}^{\ell_1 \ell_3 \ell_3}_{0 m_3 -m_3}  \langle C_{\ell_2}^{aw}\rangle  \langle C_{\ell_3}^{a'w'}\rangle \nonumber
    \\&=0 \text{ 
 unless $\ell_1=0$}\label{eq.term2_lastline} \,.
\end{align} 
Now consider the second $\langle aw \rangle \langle aw \rangle$ term in Eq.~\eqref{eq.eight_terms}:
\begin{align}
    &\frac{1}{2\ell_1+1} \sum_{m_1=-\ell_1}^{\ell_1}\sum_{\ell_2 m_2} \sum_{\ell_3 m_3}  (-1)^{m_1+m_2+m_3} \mathcal{G}^{\ell_1 \ell_2 \ell_3}_{-m_1 m_2 -m_3}  \mathcal{G}^{\ell_1 \ell_3 \ell_2}_{m_1 m_3 -m_2} \langle C_{\ell_2}^{aw'}\rangle  \langle C_{\ell_3}^{a'w}\rangle \nonumber
    \\&= \frac{1}{4\pi} \sum_{\ell_2,\ell_3} (2\ell_2+1)(2\ell_3+1) \begin{pmatrix} \ell_1&\ell_2&\ell_3 \\ 0&0&0 \end{pmatrix}^2 \langle C_{\ell_2}^{aw'}\rangle  \langle C_{\ell_3}^{a'w}\rangle \label{eq.term3_lastline} \,.
\end{align}
The first $\langle w \rangle \langle aaw \rangle_c$ term in Eq.~\eqref{eq.eight_terms} becomes the following:
\begin{align}
    &\frac{1}{2\ell_1+1} \sum_{m_1=-\ell_1}^{\ell_1}\sum_{\ell_2 m_2}\sum_{\ell_3 m_3} \sum_{\ell_4 m_4} \sum_{\ell_5 m_5}   (-1)^{m_1}\mathcal{G}^{\ell_1 \ell_2 \ell_4}_{-m_1 m_2 m_4} \mathcal{G}^{\ell_1 \ell_3 \ell_5}_{m_1 m_3 m_5} \mathcal{G}^{\ell_2 \ell_3 \ell_5}_{m_2 m_3 m_5} \langle w_{\ell_4 m_4} \rangle  \langle b_{\ell_2 \ell_3 \ell_5}^{aa'w'} \rangle \nonumber
    \\&=\frac{1}{(4\pi)^{3/2}} \sum_{\ell_2,\ell_3} (2\ell_2+1)(2\ell_3+1) \begin{pmatrix} \ell_1&\ell_2&\ell_3 \\ 0&0&0 \end{pmatrix}^2 \langle w_{00} \rangle  \langle b_{\ell_1 \ell_2 \ell_3}^{aa'w'} \rangle \label{eq.term4_lastline} \,.
\end{align}
 The other $\langle w \rangle \langle aaw \rangle_c$ term and the $\langle a \rangle \langle waw \rangle_c$ terms in Eq.~\eqref{eq.eight_terms} are similarly simplified. Finally, consider the $\langle aaww \rangle_c$ term in Eq.~\eqref{eq.eight_terms}. From Eq.~\eqref{eq.rho_expectation}, this term is simply
\begin{equation}
    \frac{1}{2\ell_1+1} \sum_{\ell_2 \ell_3 \ell_4 \ell_5} \langle \hat{\rho}[awa'w']^{\ell_2 \ell_4}_{\ell_3 \ell_5}(\ell_1) \rangle \,.
\end{equation}
Combining the results for all the terms yields
\begin{align}
    \langle C_{\ell_1}^{\tilde{a} \tilde{a}'} \rangle &= \frac{1}{4\pi} \sum_{\ell_2,\ell_3} (2\ell_2+1)(2\ell_3+1) \begin{pmatrix} \ell_1&\ell_2&\ell_3 \\ 0&0&0 \end{pmatrix}^2 \Bigg[  \langle C_{\ell_2}^{aa'}\rangle
    \langle C_{\ell_3}^{ww'}\rangle +  \langle C_{\ell_2}^{aw'}\rangle  \langle C_{\ell_3}^{a'w}\rangle   \nonumber
    \\&\qquad + \frac{\langle w_{00} \rangle}{2\sqrt{\pi}}  \langle b_{\ell_1 \ell_2 \ell_3}^{aa'w'} \rangle + \frac{\langle w'_{00} \rangle}{2\sqrt{\pi}}  \langle b_{\ell_1 \ell_2 \ell_3}^{aa'w} \rangle + \frac{\langle a_{00} \rangle}{2\sqrt{\pi}}  \langle b_{\ell_1 \ell_2 \ell_3}^{wa'w'} \rangle + \frac{\langle a'_{00} \rangle}{2\sqrt{\pi}}  \langle b_{\ell_1 \ell_2 \ell_3}^{waw'} \rangle \Bigg] 
    + \frac{1}{2\ell_1+1} \sum_{\ell_2 \ell_3 \ell_4 \ell_5} \langle \hat{\rho}[awa'w']^{\ell_2 \ell_4}_{\ell_3 \ell_5}(\ell_1) \rangle \, .
\end{align}

\end{appendices}

\bibliography{refs}

\begin{thebibliography}{68}
\expandafter\ifx\csname natexlab\endcsname\relax\def\natexlab#1{#1}\fi
\expandafter\ifx\csname bibnamefont\endcsname\relax
  \def\bibnamefont#1{#1}\fi
\expandafter\ifx\csname bibfnamefont\endcsname\relax
  \def\bibfnamefont#1{#1}\fi
\expandafter\ifx\csname citenamefont\endcsname\relax
  \def\citenamefont#1{#1}\fi
\expandafter\ifx\csname url\endcsname\relax
  \def\url#1{\texttt{#1}}\fi
\expandafter\ifx\csname urlprefix\endcsname\relax\def\urlprefix{URL }\fi
\providecommand{\bibinfo}[2]{#2}
\providecommand{\eprint}[2][]{\url{#2}}

\bibitem[{\citenamefont{Aghanim
  et~al.}(2020{\natexlab{a}})\citenamefont{Aghanim, Akrami, Ashdown, Aumont,
  Baccigalupi, Ballardini, Banday, Barreiro, Bartolo, Basak
  et~al.}}]{Planck2018}
\bibinfo{author}{\bibfnamefont{N.}~\bibnamefont{Aghanim}},
  \bibinfo{author}{\bibfnamefont{Y.}~\bibnamefont{Akrami}},
  \bibinfo{author}{\bibfnamefont{M.}~\bibnamefont{Ashdown}},
  \bibinfo{author}{\bibfnamefont{J.}~\bibnamefont{Aumont}},
  \bibinfo{author}{\bibfnamefont{C.}~\bibnamefont{Baccigalupi}},
  \bibinfo{author}{\bibfnamefont{M.}~\bibnamefont{Ballardini}},
  \bibinfo{author}{\bibfnamefont{A.~J.} \bibnamefont{Banday}},
  \bibinfo{author}{\bibfnamefont{R.~B.} \bibnamefont{Barreiro}},
  \bibinfo{author}{\bibfnamefont{N.}~\bibnamefont{Bartolo}},
  \bibinfo{author}{\bibfnamefont{S.}~\bibnamefont{Basak}},
  \bibnamefont{et~al.}, \bibinfo{journal}{Astronomy \& Astrophysics}
  \textbf{\bibinfo{volume}{641}}, \bibinfo{pages}{A5}
  (\bibinfo{year}{2020}{\natexlab{a}}), ISSN \bibinfo{issn}{1432-0746},
  \urlprefix\url{https://arxiv.org/abs/1907.12875}.

\bibitem[{\citenamefont{Coulton et~al.}(2023{\natexlab{a}})}]{Coulton2023ACT}
\bibinfo{author}{\bibfnamefont{W.}~\bibnamefont{Coulton}} \bibnamefont{et~al.}
  (\bibinfo{collaboration}{ACT}) (\bibinfo{year}{2023}{\natexlab{a}}),
  \eprint{2307.01258}, \urlprefix\url{https://arxiv.org/pdf/2307.01258.pdf}.

\bibitem[{\citenamefont{Bleem et~al.}(2022{\natexlab{a}})}]{Bleem2022SPT}
\bibinfo{author}{\bibfnamefont{L.~E.} \bibnamefont{Bleem}} \bibnamefont{et~al.}
  (\bibinfo{collaboration}{SPT-SZ}), \bibinfo{journal}{Astrophys. J. Supp.}
  \textbf{\bibinfo{volume}{258}}, \bibinfo{pages}{36}
  (\bibinfo{year}{2022}{\natexlab{a}}), \eprint{2102.05033},
  \urlprefix\url{https://arxiv.org/pdf/2102.05033.pdf}.

\bibitem[{\citenamefont{{Mart{\'\i}nez-Gonz{\'a}lez}
  et~al.}(2003)\citenamefont{{Mart{\'\i}nez-Gonz{\'a}lez}, {Diego}, {Vielva},
  and {Silk}}}]{2003MartinezGonzalez}
\bibinfo{author}{\bibfnamefont{E.}~\bibnamefont{{Mart{\'\i}nez-Gonz{\'a}lez}}},
  \bibinfo{author}{\bibfnamefont{J.~M.} \bibnamefont{{Diego}}},
  \bibinfo{author}{\bibfnamefont{P.}~\bibnamefont{{Vielva}}}, \bibnamefont{and}
  \bibinfo{author}{\bibfnamefont{J.}~\bibnamefont{{Silk}}},
  \bibinfo{journal}{\mnras} \textbf{\bibinfo{volume}{345}},
  \bibinfo{pages}{1101} (\bibinfo{year}{2003}), \eprint{astro-ph/0302094},
  \urlprefix\url{https://browse.arxiv.org/pdf/astro-ph/0302094.pdf}.

\bibitem[{\citenamefont{{Leach} et~al.}(2008)\citenamefont{{Leach}, {Cardoso},
  {Baccigalupi}, {Barreiro}, {Betoule}, {Bobin}, {Bonaldi}, {Delabrouille}, {de
  Zotti}, {Dickinson} et~al.}}]{2008Leach}
\bibinfo{author}{\bibfnamefont{S.~M.} \bibnamefont{{Leach}}},
  \bibinfo{author}{\bibfnamefont{J.~F.} \bibnamefont{{Cardoso}}},
  \bibinfo{author}{\bibfnamefont{C.}~\bibnamefont{{Baccigalupi}}},
  \bibinfo{author}{\bibfnamefont{R.~B.} \bibnamefont{{Barreiro}}},
  \bibinfo{author}{\bibfnamefont{M.}~\bibnamefont{{Betoule}}},
  \bibinfo{author}{\bibfnamefont{J.}~\bibnamefont{{Bobin}}},
  \bibinfo{author}{\bibfnamefont{A.}~\bibnamefont{{Bonaldi}}},
  \bibinfo{author}{\bibfnamefont{J.}~\bibnamefont{{Delabrouille}}},
  \bibinfo{author}{\bibfnamefont{G.}~\bibnamefont{{de Zotti}}},
  \bibinfo{author}{\bibfnamefont{C.}~\bibnamefont{{Dickinson}}},
  \bibnamefont{et~al.}, \bibinfo{journal}{\aap} \textbf{\bibinfo{volume}{491}},
  \bibinfo{pages}{597} (\bibinfo{year}{2008}), \eprint{0805.0269},
  \urlprefix\url{https://browse.arxiv.org/pdf/0805.0269.pdf}.

\bibitem[{\citenamefont{{Fern{\'a}ndez-Cobos}
  et~al.}(2012)\citenamefont{{Fern{\'a}ndez-Cobos}, {Vielva}, {Barreiro}, and
  {Mart{\'\i}nez-Gonz{\'a}lez}}}]{2012FernandezCobos}
\bibinfo{author}{\bibfnamefont{R.}~\bibnamefont{{Fern{\'a}ndez-Cobos}}},
  \bibinfo{author}{\bibfnamefont{P.}~\bibnamefont{{Vielva}}},
  \bibinfo{author}{\bibfnamefont{R.~B.} \bibnamefont{{Barreiro}}},
  \bibnamefont{and}
  \bibinfo{author}{\bibfnamefont{E.}~\bibnamefont{{Mart{\'\i}nez-Gonz{\'a}lez}}},
  \bibinfo{journal}{\mnras} \textbf{\bibinfo{volume}{420}},
  \bibinfo{pages}{2162} (\bibinfo{year}{2012}), \eprint{1106.2016},
  \urlprefix\url{https://browse.arxiv.org/pdf/1106.2016.pdf}.

\bibitem[{\citenamefont{{Delabrouille}
  et~al.}(2003)\citenamefont{{Delabrouille}, {Cardoso}, and
  {Patanchon}}}]{2003Delabrouille}
\bibinfo{author}{\bibfnamefont{J.}~\bibnamefont{{Delabrouille}}},
  \bibinfo{author}{\bibfnamefont{J.~F.} \bibnamefont{{Cardoso}}},
  \bibnamefont{and}
  \bibinfo{author}{\bibfnamefont{G.}~\bibnamefont{{Patanchon}}},
  \bibinfo{journal}{\mnras} \textbf{\bibinfo{volume}{346}},
  \bibinfo{pages}{1089} (\bibinfo{year}{2003}), \eprint{astro-ph/0211504},
  \urlprefix\url{https://browse.arxiv.org/pdf/astro-ph/0211504.pdf}.

\bibitem[{\citenamefont{{Cardoso} et~al.}(2008)\citenamefont{{Cardoso}, {Le
  Jeune}, {Delabrouille}, {Betoule}, and {Patanchon}}}]{2008Cardoso}
\bibinfo{author}{\bibfnamefont{J.-F.} \bibnamefont{{Cardoso}}},
  \bibinfo{author}{\bibfnamefont{M.}~\bibnamefont{{Le Jeune}}},
  \bibinfo{author}{\bibfnamefont{J.}~\bibnamefont{{Delabrouille}}},
  \bibinfo{author}{\bibfnamefont{M.}~\bibnamefont{{Betoule}}},
  \bibnamefont{and}
  \bibinfo{author}{\bibfnamefont{G.}~\bibnamefont{{Patanchon}}},
  \bibinfo{journal}{IEEE Journal of Selected Topics in Signal Processing}
  \textbf{\bibinfo{volume}{2}}, \bibinfo{pages}{735} (\bibinfo{year}{2008}),
  \urlprefix\url{https://ieeexplore.ieee.org/document/4703509}.

\bibitem[{\citenamefont{{Eriksen} et~al.}(2006)\citenamefont{{Eriksen},
  {Dickinson}, {Lawrence}, {Baccigalupi}, {Banday}, {G{\'o}rski}, {Hansen},
  {Lilje}, {Pierpaoli}, {Seiffert} et~al.}}]{2006Eriksen}
\bibinfo{author}{\bibfnamefont{H.~K.} \bibnamefont{{Eriksen}}},
  \bibinfo{author}{\bibfnamefont{C.}~\bibnamefont{{Dickinson}}},
  \bibinfo{author}{\bibfnamefont{C.~R.} \bibnamefont{{Lawrence}}},
  \bibinfo{author}{\bibfnamefont{C.}~\bibnamefont{{Baccigalupi}}},
  \bibinfo{author}{\bibfnamefont{A.~J.} \bibnamefont{{Banday}}},
  \bibinfo{author}{\bibfnamefont{K.~M.} \bibnamefont{{G{\'o}rski}}},
  \bibinfo{author}{\bibfnamefont{F.~K.} \bibnamefont{{Hansen}}},
  \bibinfo{author}{\bibfnamefont{P.~B.} \bibnamefont{{Lilje}}},
  \bibinfo{author}{\bibfnamefont{E.}~\bibnamefont{{Pierpaoli}}},
  \bibinfo{author}{\bibfnamefont{M.~D.} \bibnamefont{{Seiffert}}},
  \bibnamefont{et~al.}, \bibinfo{journal}{\apj} \textbf{\bibinfo{volume}{641}},
  \bibinfo{pages}{665} (\bibinfo{year}{2006}), \eprint{astro-ph/0508268},
  \urlprefix\url{https://browse.arxiv.org/pdf/astro-ph/0508268.pdf}.

\bibitem[{\citenamefont{{Eriksen} et~al.}(2008)\citenamefont{{Eriksen},
  {Jewell}, {Dickinson}, {Banday}, {G{\'o}rski}, and {Lawrence}}}]{2008Eriksen}
\bibinfo{author}{\bibfnamefont{H.~K.} \bibnamefont{{Eriksen}}},
  \bibinfo{author}{\bibfnamefont{J.~B.} \bibnamefont{{Jewell}}},
  \bibinfo{author}{\bibfnamefont{C.}~\bibnamefont{{Dickinson}}},
  \bibinfo{author}{\bibfnamefont{A.~J.} \bibnamefont{{Banday}}},
  \bibinfo{author}{\bibfnamefont{K.~M.} \bibnamefont{{G{\'o}rski}}},
  \bibnamefont{and} \bibinfo{author}{\bibfnamefont{C.~R.}
  \bibnamefont{{Lawrence}}}, \bibinfo{journal}{\apj}
  \textbf{\bibinfo{volume}{676}}, \bibinfo{pages}{10} (\bibinfo{year}{2008}),
  \eprint{0709.1058},
  \urlprefix\url{https://browse.arxiv.org/pdf/0709.1058.pdf}.

\bibitem[{\citenamefont{{Bennett} et~al.}(2003)\citenamefont{{Bennett}, {Hill},
  {Hinshaw}, {Nolta}, {Odegard}, {Page}, {Spergel}, {Weiland}, {Wright},
  {Halpern} et~al.}}]{Bennett2003}
\bibinfo{author}{\bibfnamefont{C.~L.} \bibnamefont{{Bennett}}},
  \bibinfo{author}{\bibfnamefont{R.~S.} \bibnamefont{{Hill}}},
  \bibinfo{author}{\bibfnamefont{G.}~\bibnamefont{{Hinshaw}}},
  \bibinfo{author}{\bibfnamefont{M.~R.} \bibnamefont{{Nolta}}},
  \bibinfo{author}{\bibfnamefont{N.}~\bibnamefont{{Odegard}}},
  \bibinfo{author}{\bibfnamefont{L.}~\bibnamefont{{Page}}},
  \bibinfo{author}{\bibfnamefont{D.~N.} \bibnamefont{{Spergel}}},
  \bibinfo{author}{\bibfnamefont{J.~L.} \bibnamefont{{Weiland}}},
  \bibinfo{author}{\bibfnamefont{E.~L.} \bibnamefont{{Wright}}},
  \bibinfo{author}{\bibfnamefont{M.}~\bibnamefont{{Halpern}}},
  \bibnamefont{et~al.}, \bibinfo{journal}{\apjs}
  \textbf{\bibinfo{volume}{148}}, \bibinfo{pages}{97} (\bibinfo{year}{2003}),
  \eprint{astro-ph/0302208},
  \urlprefix\url{https://arxiv.org/abs/astro-ph/0302208}.

\bibitem[{\citenamefont{{Tegmark} et~al.}(2003)\citenamefont{{Tegmark}, {de
  Oliveira-Costa}, and {Hamilton}}}]{Tegmark2003}
\bibinfo{author}{\bibfnamefont{M.}~\bibnamefont{{Tegmark}}},
  \bibinfo{author}{\bibfnamefont{A.}~\bibnamefont{{de Oliveira-Costa}}},
  \bibnamefont{and} \bibinfo{author}{\bibfnamefont{A.~J.}
  \bibnamefont{{Hamilton}}}, \bibinfo{journal}{\prd}
  \textbf{\bibinfo{volume}{68}}, \bibinfo{eid}{123523} (\bibinfo{year}{2003}),
  \eprint{astro-ph/0302496},
  \urlprefix\url{https://arxiv.org/abs/astro-ph/0302496}.

\bibitem[{\citenamefont{{Eriksen} et~al.}(2004)\citenamefont{{Eriksen},
  {Banday}, {G{\'o}rski}, and {Lilje}}}]{Eriksen2004}
\bibinfo{author}{\bibfnamefont{H.~K.} \bibnamefont{{Eriksen}}},
  \bibinfo{author}{\bibfnamefont{A.~J.} \bibnamefont{{Banday}}},
  \bibinfo{author}{\bibfnamefont{K.~M.} \bibnamefont{{G{\'o}rski}}},
  \bibnamefont{and} \bibinfo{author}{\bibfnamefont{P.~B.}
  \bibnamefont{{Lilje}}}, \bibinfo{journal}{\apj}
  \textbf{\bibinfo{volume}{612}}, \bibinfo{pages}{633} (\bibinfo{year}{2004}),
  \eprint{astro-ph/0403098},
  \urlprefix\url{https://arxiv.org/abs/astro-ph/0403098}.

\bibitem[{\citenamefont{{Delabrouille}
  et~al.}(2009)\citenamefont{{Delabrouille}, {Cardoso}, {Le Jeune}, {Betoule},
  {Fay}, and {Guilloux}}}]{Delabrouille2009}
\bibinfo{author}{\bibfnamefont{J.}~\bibnamefont{{Delabrouille}}},
  \bibinfo{author}{\bibfnamefont{J.~F.} \bibnamefont{{Cardoso}}},
  \bibinfo{author}{\bibfnamefont{M.}~\bibnamefont{{Le Jeune}}},
  \bibinfo{author}{\bibfnamefont{M.}~\bibnamefont{{Betoule}}},
  \bibinfo{author}{\bibfnamefont{G.}~\bibnamefont{{Fay}}}, \bibnamefont{and}
  \bibinfo{author}{\bibfnamefont{F.}~\bibnamefont{{Guilloux}}},
  \bibinfo{journal}{\aap} \textbf{\bibinfo{volume}{493}}, \bibinfo{pages}{835}
  (\bibinfo{year}{2009}), \eprint{0807.0773},
  \urlprefix\url{https://arxiv.org/abs/0807.0773}.

\bibitem[{\citenamefont{McCarthy and
  Hill}(2023{\natexlab{a}})}]{McCarthy:2023hpa}
\bibinfo{author}{\bibfnamefont{F.}~\bibnamefont{McCarthy}} \bibnamefont{and}
  \bibinfo{author}{\bibfnamefont{J.~C.} \bibnamefont{Hill}}
  (\bibinfo{year}{2023}{\natexlab{a}}), \eprint{2307.01043},
  \urlprefix\url{https://arxiv.org/pdf/2307.01043.pdf}.

\bibitem[{\citenamefont{Chandran et~al.}(2023)\citenamefont{Chandran,
  Remazeilles, and Barreiro}}]{Chandran:2023akr}
\bibinfo{author}{\bibfnamefont{J.}~\bibnamefont{Chandran}},
  \bibinfo{author}{\bibfnamefont{M.}~\bibnamefont{Remazeilles}},
  \bibnamefont{and} \bibinfo{author}{\bibfnamefont{R.~B.}
  \bibnamefont{Barreiro}} (\bibinfo{year}{2023}), \eprint{2305.10193},
  \urlprefix\url{https://arxiv.org/pdf/2305.10193.pdf}.

\bibitem[{\citenamefont{Coulton et~al.}(2023{\natexlab{b}})}]{ACT:2023wcq}
\bibinfo{author}{\bibfnamefont{W.}~\bibnamefont{Coulton}} \bibnamefont{et~al.}
  (\bibinfo{collaboration}{ACT}) (\bibinfo{year}{2023}{\natexlab{b}}),
  \eprint{2307.01258}, \urlprefix\url{https://arxiv.org/pdf/2307.01258.pdf}.

\bibitem[{\citenamefont{Bleem et~al.}(2022{\natexlab{b}})}]{SPT-SZ:2021gsa}
\bibinfo{author}{\bibfnamefont{L.~E.} \bibnamefont{Bleem}} \bibnamefont{et~al.}
  (\bibinfo{collaboration}{SPT-SZ}), \bibinfo{journal}{Astrophys. J. Supp.}
  \textbf{\bibinfo{volume}{258}}, \bibinfo{pages}{36}
  (\bibinfo{year}{2022}{\natexlab{b}}), \eprint{2102.05033},
  \urlprefix\url{https://arxiv.org/pdf/2102.05033.pdf}.

\bibitem[{\citenamefont{Remazeilles et~al.}(2013)\citenamefont{Remazeilles,
  Aghanim, and Douspis}}]{Remazeilles:2012pn}
\bibinfo{author}{\bibfnamefont{M.}~\bibnamefont{Remazeilles}},
  \bibinfo{author}{\bibfnamefont{N.}~\bibnamefont{Aghanim}}, \bibnamefont{and}
  \bibinfo{author}{\bibfnamefont{M.}~\bibnamefont{Douspis}},
  \bibinfo{journal}{Mon. Not. Roy. Astron. Soc.}
  \textbf{\bibinfo{volume}{430}}, \bibinfo{pages}{370} (\bibinfo{year}{2013}),
  \eprint{1207.4683}, \urlprefix\url{https://arxiv.org/pdf/1207.4683.pdf}.

\bibitem[{\citenamefont{Wolz et~al.}(2023)}]{Wolz:2023lzb}
\bibinfo{author}{\bibfnamefont{K.}~\bibnamefont{Wolz}} \bibnamefont{et~al.}
  (\bibinfo{year}{2023}), \eprint{2302.04276},
  \urlprefix\url{https://arxiv.org/pdf/2302.04276.pdf}.

\bibitem[{\citenamefont{Carones et~al.}(2022)\citenamefont{Carones, Migliaccio,
  Puglisi, Baccigalupi, Marinucci, Vittorio, and Poletti}}]{Carones:2022xzs}
\bibinfo{author}{\bibfnamefont{A.}~\bibnamefont{Carones}},
  \bibinfo{author}{\bibfnamefont{M.}~\bibnamefont{Migliaccio}},
  \bibinfo{author}{\bibfnamefont{G.}~\bibnamefont{Puglisi}},
  \bibinfo{author}{\bibfnamefont{C.}~\bibnamefont{Baccigalupi}},
  \bibinfo{author}{\bibfnamefont{D.}~\bibnamefont{Marinucci}},
  \bibinfo{author}{\bibfnamefont{N.}~\bibnamefont{Vittorio}}, \bibnamefont{and}
  \bibinfo{author}{\bibfnamefont{D.}~\bibnamefont{Poletti}}
  (\bibinfo{year}{2022}), \eprint{2212.04456},
  \urlprefix\url{https://arxiv.org/pdf/2212.04456.pdf}.

\bibitem[{\citenamefont{Abazajian et~al.}(2019)}]{Abazajian:2019eic}
\bibinfo{author}{\bibfnamefont{K.}~\bibnamefont{Abazajian}}
  \bibnamefont{et~al.} (\bibinfo{year}{2019}), \eprint{1907.04473},
  \urlprefix\url{https://arxiv.org/pdf/1907.04473.pdf}.

\bibitem[{\citenamefont{{Planck Collaboration}
  et~al.}(2016{\natexlab{a}})\citenamefont{{Planck Collaboration}, {Aghanim},
  {Arnaud}, {Ashdown}, {Aumont}, {Baccigalupi}, {Banday}, {Barreiro},
  {Bartlett}, {Bartolo} et~al.}}]{Planck2015ymap}
\bibinfo{author}{\bibnamefont{{Planck Collaboration}}},
  \bibinfo{author}{\bibfnamefont{N.}~\bibnamefont{{Aghanim}}},
  \bibinfo{author}{\bibfnamefont{M.}~\bibnamefont{{Arnaud}}},
  \bibinfo{author}{\bibfnamefont{M.}~\bibnamefont{{Ashdown}}},
  \bibinfo{author}{\bibfnamefont{J.}~\bibnamefont{{Aumont}}},
  \bibinfo{author}{\bibfnamefont{C.}~\bibnamefont{{Baccigalupi}}},
  \bibinfo{author}{\bibfnamefont{A.~J.} \bibnamefont{{Banday}}},
  \bibinfo{author}{\bibfnamefont{R.~B.} \bibnamefont{{Barreiro}}},
  \bibinfo{author}{\bibfnamefont{J.~G.} \bibnamefont{{Bartlett}}},
  \bibinfo{author}{\bibfnamefont{N.}~\bibnamefont{{Bartolo}}},
  \bibnamefont{et~al.}, \bibinfo{journal}{\aap} \textbf{\bibinfo{volume}{594}},
  \bibinfo{eid}{A22} (\bibinfo{year}{2016}{\natexlab{a}}), \eprint{1502.01596},
  \urlprefix\url{https://arxiv.org/pdf/1502.01596.pdf}.

\bibitem[{\citenamefont{{Bolliet} et~al.}(2018)\citenamefont{{Bolliet},
  {Comis}, {Komatsu}, and {Mac{\'\i}as-P{\'e}rez}}}]{Bolliet2018}
\bibinfo{author}{\bibfnamefont{B.}~\bibnamefont{{Bolliet}}},
  \bibinfo{author}{\bibfnamefont{B.}~\bibnamefont{{Comis}}},
  \bibinfo{author}{\bibfnamefont{E.}~\bibnamefont{{Komatsu}}},
  \bibnamefont{and} \bibinfo{author}{\bibfnamefont{J.~F.}
  \bibnamefont{{Mac{\'\i}as-P{\'e}rez}}}, \bibinfo{journal}{\mnras}
  \textbf{\bibinfo{volume}{477}}, \bibinfo{pages}{4957} (\bibinfo{year}{2018}),
  \eprint{1712.00788}, \urlprefix\url{https://arxiv.org/pdf/1712.00788.pdf}.

\bibitem[{\citenamefont{Rotti et~al.}(2021)\citenamefont{Rotti, Bolliet,
  Chluba, and Remazeilles}}]{Rotti:2020rdl}
\bibinfo{author}{\bibfnamefont{A.}~\bibnamefont{Rotti}},
  \bibinfo{author}{\bibfnamefont{B.}~\bibnamefont{Bolliet}},
  \bibinfo{author}{\bibfnamefont{J.}~\bibnamefont{Chluba}}, \bibnamefont{and}
  \bibinfo{author}{\bibfnamefont{M.}~\bibnamefont{Remazeilles}},
  \bibinfo{journal}{Mon. Not. Roy. Astron. Soc.}
  \textbf{\bibinfo{volume}{503}}, \bibinfo{pages}{5310} (\bibinfo{year}{2021}),
  \eprint{2010.07797}, \urlprefix\url{https://arxiv.org/pdf/2010.07797.pdf}.

\bibitem[{\citenamefont{{Planck Collaboration}
  et~al.}(2016{\natexlab{b}})\citenamefont{{Planck Collaboration}, {Adam},
  {Ade}, {Aghanim}, {Arnaud}, {Ashdown}, {Aumont}, {Baccigalupi}, {Banday},
  {Barreiro} et~al.}}]{Planck2015compsep}
\bibinfo{author}{\bibnamefont{{Planck Collaboration}}},
  \bibinfo{author}{\bibfnamefont{R.}~\bibnamefont{{Adam}}},
  \bibinfo{author}{\bibfnamefont{P.~A.~R.} \bibnamefont{{Ade}}},
  \bibinfo{author}{\bibfnamefont{N.}~\bibnamefont{{Aghanim}}},
  \bibinfo{author}{\bibfnamefont{M.}~\bibnamefont{{Arnaud}}},
  \bibinfo{author}{\bibfnamefont{M.}~\bibnamefont{{Ashdown}}},
  \bibinfo{author}{\bibfnamefont{J.}~\bibnamefont{{Aumont}}},
  \bibinfo{author}{\bibfnamefont{C.}~\bibnamefont{{Baccigalupi}}},
  \bibinfo{author}{\bibfnamefont{A.~J.} \bibnamefont{{Banday}}},
  \bibinfo{author}{\bibfnamefont{R.~B.} \bibnamefont{{Barreiro}}},
  \bibnamefont{et~al.}, \bibinfo{journal}{\aap} \textbf{\bibinfo{volume}{594}},
  \bibinfo{eid}{A9} (\bibinfo{year}{2016}{\natexlab{b}}), \eprint{1502.05956},
  \urlprefix\url{https://arxiv.org/abs/1502.05956}.

\bibitem[{\citenamefont{Dunkley et~al.}(2013)\citenamefont{Dunkley, Calabrese,
  Sievers, Addison, Battaglia, Battistelli, Bond, Das, Devlin, Dünner
  et~al.}}]{Dunkley}
\bibinfo{author}{\bibfnamefont{J.}~\bibnamefont{Dunkley}},
  \bibinfo{author}{\bibfnamefont{E.}~\bibnamefont{Calabrese}},
  \bibinfo{author}{\bibfnamefont{J.}~\bibnamefont{Sievers}},
  \bibinfo{author}{\bibfnamefont{G.}~\bibnamefont{Addison}},
  \bibinfo{author}{\bibfnamefont{N.}~\bibnamefont{Battaglia}},
  \bibinfo{author}{\bibfnamefont{E.}~\bibnamefont{Battistelli}},
  \bibinfo{author}{\bibfnamefont{J.}~\bibnamefont{Bond}},
  \bibinfo{author}{\bibfnamefont{S.}~\bibnamefont{Das}},
  \bibinfo{author}{\bibfnamefont{M.}~\bibnamefont{Devlin}},
  \bibinfo{author}{\bibfnamefont{R.}~\bibnamefont{Dünner}},
  \bibnamefont{et~al.}, \bibinfo{journal}{Journal of Cosmology and
  Astroparticle Physics} \textbf{\bibinfo{volume}{2013}},
  \bibinfo{pages}{025–025} (\bibinfo{year}{2013}), ISSN
  \bibinfo{issn}{1475-7516}, \urlprefix\url{https://arxiv.org/abs/1301.0776}.

\bibitem[{\citenamefont{Aiola et~al.}(2020)\citenamefont{Aiola, Calabrese,
  Maurin, Naess, Schmitt, Abitbol, Addison, Ade, Alonso, Amiri
  et~al.}}]{ACT2020}
\bibinfo{author}{\bibfnamefont{S.}~\bibnamefont{Aiola}},
  \bibinfo{author}{\bibfnamefont{E.}~\bibnamefont{Calabrese}},
  \bibinfo{author}{\bibfnamefont{L.}~\bibnamefont{Maurin}},
  \bibinfo{author}{\bibfnamefont{S.}~\bibnamefont{Naess}},
  \bibinfo{author}{\bibfnamefont{B.~L.} \bibnamefont{Schmitt}},
  \bibinfo{author}{\bibfnamefont{M.~H.} \bibnamefont{Abitbol}},
  \bibinfo{author}{\bibfnamefont{G.~E.} \bibnamefont{Addison}},
  \bibinfo{author}{\bibfnamefont{P.~A.~R.} \bibnamefont{Ade}},
  \bibinfo{author}{\bibfnamefont{D.}~\bibnamefont{Alonso}},
  \bibinfo{author}{\bibfnamefont{M.}~\bibnamefont{Amiri}},
  \bibnamefont{et~al.}, \bibinfo{journal}{Journal of Cosmology and
  Astroparticle Physics} \textbf{\bibinfo{volume}{2020}},
  \bibinfo{pages}{047–047} (\bibinfo{year}{2020}), ISSN
  \bibinfo{issn}{1475-7516}, \urlprefix\url{https://arxiv.org/abs/2007.07288}.

\bibitem[{\citenamefont{Dutcher et~al.}(2021)\citenamefont{Dutcher, Balkenhol,
  Ade, Ahmed, Anderes, Anderson, Archipley, Avva, Aylor, Barry et~al.}}]{SPT}
\bibinfo{author}{\bibfnamefont{D.}~\bibnamefont{Dutcher}},
  \bibinfo{author}{\bibfnamefont{L.}~\bibnamefont{Balkenhol}},
  \bibinfo{author}{\bibfnamefont{P.}~\bibnamefont{Ade}},
  \bibinfo{author}{\bibfnamefont{Z.}~\bibnamefont{Ahmed}},
  \bibinfo{author}{\bibfnamefont{E.}~\bibnamefont{Anderes}},
  \bibinfo{author}{\bibfnamefont{A.}~\bibnamefont{Anderson}},
  \bibinfo{author}{\bibfnamefont{M.}~\bibnamefont{Archipley}},
  \bibinfo{author}{\bibfnamefont{J.}~\bibnamefont{Avva}},
  \bibinfo{author}{\bibfnamefont{K.}~\bibnamefont{Aylor}},
  \bibinfo{author}{\bibfnamefont{P.}~\bibnamefont{Barry}},
  \bibnamefont{et~al.}, \bibinfo{journal}{Physical Review D}
  \textbf{\bibinfo{volume}{104}} (\bibinfo{year}{2021}), ISSN
  \bibinfo{issn}{2470-0029}, \urlprefix\url{https://arxiv.org/abs/2101.01684}.

\bibitem[{\citenamefont{{Chen} and {Wright}}(2009)}]{2009Chen}
\bibinfo{author}{\bibfnamefont{X.}~\bibnamefont{{Chen}}} \bibnamefont{and}
  \bibinfo{author}{\bibfnamefont{E.~L.} \bibnamefont{{Wright}}},
  \bibinfo{journal}{\apj} \textbf{\bibinfo{volume}{694}}, \bibinfo{pages}{222}
  (\bibinfo{year}{2009}), \eprint{0809.4025},
  \urlprefix\url{https://arxiv.org/pdf/0809.4025.pdf}.

\bibitem[{\citenamefont{{Remazeilles} et~al.}(2011)\citenamefont{{Remazeilles},
  {Delabrouille}, and {Cardoso}}}]{2011Remazeilles}
\bibinfo{author}{\bibfnamefont{M.}~\bibnamefont{{Remazeilles}}},
  \bibinfo{author}{\bibfnamefont{J.}~\bibnamefont{{Delabrouille}}},
  \bibnamefont{and} \bibinfo{author}{\bibfnamefont{J.-F.}
  \bibnamefont{{Cardoso}}}, \bibinfo{journal}{\mnras}
  \textbf{\bibinfo{volume}{410}}, \bibinfo{pages}{2481} (\bibinfo{year}{2011}),
  \eprint{1006.5599}, \urlprefix\url{https://arxiv.org/pdf/1006.5599.pdf}.

\bibitem[{\citenamefont{{Chluba} et~al.}(2017)\citenamefont{{Chluba}, {Hill},
  and {Abitbol}}}]{2017Chluba}
\bibinfo{author}{\bibfnamefont{J.}~\bibnamefont{{Chluba}}},
  \bibinfo{author}{\bibfnamefont{J.~C.} \bibnamefont{{Hill}}},
  \bibnamefont{and} \bibinfo{author}{\bibfnamefont{M.~H.}
  \bibnamefont{{Abitbol}}}, \bibinfo{journal}{\mnras}
  \textbf{\bibinfo{volume}{472}}, \bibinfo{pages}{1195} (\bibinfo{year}{2017}),
  \eprint{1701.00274}, \urlprefix\url{https://arxiv.org/pdf/1701.00274.pdf}.

\bibitem[{\citenamefont{Remazeilles et~al.}(2021)\citenamefont{Remazeilles,
  Rotti, and Chluba}}]{Remazeilles:2020rqw}
\bibinfo{author}{\bibfnamefont{M.}~\bibnamefont{Remazeilles}},
  \bibinfo{author}{\bibfnamefont{A.}~\bibnamefont{Rotti}}, \bibnamefont{and}
  \bibinfo{author}{\bibfnamefont{J.}~\bibnamefont{Chluba}},
  \bibinfo{journal}{Mon. Not. Roy. Astron. Soc.}
  \textbf{\bibinfo{volume}{503}}, \bibinfo{pages}{2478} (\bibinfo{year}{2021}),
  \eprint{2006.08628},
  \urlprefix\url{https://browse.arxiv.org/pdf/2006.08628.pdf}.

\bibitem[{\citenamefont{Kusiak et~al.}(2023)\citenamefont{Kusiak, Surrao, and
  Hill}}]{Kusiak:2023hrz}
\bibinfo{author}{\bibfnamefont{A.}~\bibnamefont{Kusiak}},
  \bibinfo{author}{\bibfnamefont{K.~M.} \bibnamefont{Surrao}},
  \bibnamefont{and} \bibinfo{author}{\bibfnamefont{J.~C.} \bibnamefont{Hill}}
  (\bibinfo{year}{2023}), \eprint{2303.08121},
  \urlprefix\url{https://browse.arxiv.org/pdf/2303.08121.pdf}.

\bibitem[{\citenamefont{Ade et~al.}(2014{\natexlab{a}})}]{Planck:2013compsep}
\bibinfo{author}{\bibfnamefont{P.~A.~R.} \bibnamefont{Ade}}
  \bibnamefont{et~al.} (\bibinfo{collaboration}{Planck}),
  \bibinfo{journal}{Astron. Astrophys.} \textbf{\bibinfo{volume}{571}},
  \bibinfo{pages}{A12} (\bibinfo{year}{2014}{\natexlab{a}}),
  \eprint{1303.5072}, \urlprefix\url{https://arxiv.org/pdf/1303.5072.pdf}.

\bibitem[{\citenamefont{Ade et~al.}(2014{\natexlab{b}})}]{Planck:2013ymap}
\bibinfo{author}{\bibfnamefont{P.~A.~R.} \bibnamefont{Ade}}
  \bibnamefont{et~al.} (\bibinfo{collaboration}{Planck}),
  \bibinfo{journal}{Astron. Astrophys.} \textbf{\bibinfo{volume}{571}},
  \bibinfo{pages}{A21} (\bibinfo{year}{2014}{\natexlab{b}}),
  \eprint{1303.5081}, \urlprefix\url{https://arxiv.org/pdf/1303.5081.pdf}.

\bibitem[{\citenamefont{McCarthy and
  Hill}(2023{\natexlab{b}})}]{McCarthy:2023cwg}
\bibinfo{author}{\bibfnamefont{F.}~\bibnamefont{McCarthy}} \bibnamefont{and}
  \bibinfo{author}{\bibfnamefont{J.~C.} \bibnamefont{Hill}}
  (\bibinfo{year}{2023}{\natexlab{b}}), \eprint{2308.16260},
  \urlprefix\url{https://arxiv.org/pdf/2308.16260.pdf}.

\bibitem[{\citenamefont{Narcowich et~al.}(2006)\citenamefont{Narcowich,
  Petrushev, and Ward}}]{Narcowich2006}
\bibinfo{author}{\bibfnamefont{F.~J.} \bibnamefont{Narcowich}},
  \bibinfo{author}{\bibfnamefont{P.}~\bibnamefont{Petrushev}},
  \bibnamefont{and} \bibinfo{author}{\bibfnamefont{J.~D.} \bibnamefont{Ward}},
  \bibinfo{journal}{SIAM Journal on Mathematical Analysis}
  \textbf{\bibinfo{volume}{38}}, \bibinfo{pages}{574} (\bibinfo{year}{2006}),
  \urlprefix\url{https://www.di.ens.fr/~mallat/papiers/040614359.pdf}.

\bibitem[{\citenamefont{Marinucci et~al.}(2007)\citenamefont{Marinucci,
  Pietrobon, Balbi, Baldi, Cabella, Kerkyacharian, Natoli, Picard, and
  Vittorio}}]{Marinucci_2007}
\bibinfo{author}{\bibfnamefont{D.}~\bibnamefont{Marinucci}},
  \bibinfo{author}{\bibfnamefont{D.}~\bibnamefont{Pietrobon}},
  \bibinfo{author}{\bibfnamefont{A.}~\bibnamefont{Balbi}},
  \bibinfo{author}{\bibfnamefont{P.}~\bibnamefont{Baldi}},
  \bibinfo{author}{\bibfnamefont{P.}~\bibnamefont{Cabella}},
  \bibinfo{author}{\bibfnamefont{G.}~\bibnamefont{Kerkyacharian}},
  \bibinfo{author}{\bibfnamefont{P.}~\bibnamefont{Natoli}},
  \bibinfo{author}{\bibfnamefont{D.}~\bibnamefont{Picard}}, \bibnamefont{and}
  \bibinfo{author}{\bibfnamefont{N.}~\bibnamefont{Vittorio}},
  \bibinfo{journal}{Monthly Notices of the Royal Astronomical Society}
  \textbf{\bibinfo{volume}{383}}, \bibinfo{pages}{539} (\bibinfo{year}{2007}),
  \urlprefix\url{https://arxiv.org/abs/0707.0844}.

\bibitem[{\citenamefont{Guilloux et~al.}(2007)\citenamefont{Guilloux, Fay, and
  Cardoso}}]{Guilloux2008}
\bibinfo{author}{\bibfnamefont{F.}~\bibnamefont{Guilloux}},
  \bibinfo{author}{\bibfnamefont{G.}~\bibnamefont{Fay}}, \bibnamefont{and}
  \bibinfo{author}{\bibfnamefont{J.-F.} \bibnamefont{Cardoso}},
  \emph{\bibinfo{title}{Practical wavelet design on the sphere}}
  (\bibinfo{year}{2007}), \urlprefix\url{https://arxiv.org/abs/0706.2598}.

\bibitem[{\citenamefont{Komatsu and Spergel}(2001)}]{Komatsu:2001rj}
\bibinfo{author}{\bibfnamefont{E.}~\bibnamefont{Komatsu}} \bibnamefont{and}
  \bibinfo{author}{\bibfnamefont{D.~N.} \bibnamefont{Spergel}},
  \bibinfo{journal}{Phys. Rev. D} \textbf{\bibinfo{volume}{63}},
  \bibinfo{pages}{063002} (\bibinfo{year}{2001}), \eprint{astro-ph/0005036},
  \urlprefix\url{https://arxiv.org/abs/astro-ph/0005036}.

\bibitem[{\citenamefont{Fergusson and Shellard}(2009)}]{Fergusson:2008}
\bibinfo{author}{\bibfnamefont{J.~R.} \bibnamefont{Fergusson}}
  \bibnamefont{and} \bibinfo{author}{\bibfnamefont{E.~P.~S.}
  \bibnamefont{Shellard}}, \bibinfo{journal}{Phys. Rev. D}
  \textbf{\bibinfo{volume}{80}}, \bibinfo{pages}{043510}
  (\bibinfo{year}{2009}), \eprint{0812.3413},
  \urlprefix\url{https://arxiv.org/abs/0812.3413}.

\bibitem[{\citenamefont{Fergusson and Shellard}(2011)}]{fergusson2011optimal}
\bibinfo{author}{\bibfnamefont{J.}~\bibnamefont{Fergusson}} \bibnamefont{and}
  \bibinfo{author}{\bibfnamefont{E.}~\bibnamefont{Shellard}},
  \bibinfo{journal}{arXiv preprint arXiv:1105.2791}  (\bibinfo{year}{2011}),
  \urlprefix\url{https://arxiv.org/abs/1105.2791}.

\bibitem[{\citenamefont{Bucher et~al.}(2016)\citenamefont{Bucher, Racine, and
  van Tent}}]{bucher2016binned}
\bibinfo{author}{\bibfnamefont{M.}~\bibnamefont{Bucher}},
  \bibinfo{author}{\bibfnamefont{B.}~\bibnamefont{Racine}}, \bibnamefont{and}
  \bibinfo{author}{\bibfnamefont{B.}~\bibnamefont{van Tent}},
  \bibinfo{journal}{Journal of Cosmology and Astroparticle Physics}
  \textbf{\bibinfo{volume}{2016}}, \bibinfo{pages}{055} (\bibinfo{year}{2016}),
  \eprint{astro-ph/1509.08107},
  \urlprefix\url{https://arxiv.org/abs/1509.08107}.

\bibitem[{\citenamefont{Regan et~al.}(2010)\citenamefont{Regan, Shellard, and
  Fergusson}}]{Regan:2010}
\bibinfo{author}{\bibfnamefont{D.~M.} \bibnamefont{Regan}},
  \bibinfo{author}{\bibfnamefont{E.~P.~S.} \bibnamefont{Shellard}},
  \bibnamefont{and} \bibinfo{author}{\bibfnamefont{J.~R.}
  \bibnamefont{Fergusson}}, \bibinfo{journal}{Phys. Rev. D}
  \textbf{\bibinfo{volume}{82}}, \bibinfo{pages}{023520}
  (\bibinfo{year}{2010}), \eprint{1004.2915},
  \urlprefix\url{https://arxiv.org/abs/1004.2915}.

\bibitem[{\citenamefont{Fergusson et~al.}(2010)\citenamefont{Fergusson, Regan,
  and Shellard}}]{Fergusson:2010}
\bibinfo{author}{\bibfnamefont{J.~R.} \bibnamefont{Fergusson}},
  \bibinfo{author}{\bibfnamefont{D.~M.} \bibnamefont{Regan}}, \bibnamefont{and}
  \bibinfo{author}{\bibfnamefont{E.~P.~S.} \bibnamefont{Shellard}}
  (\bibinfo{year}{2010}), \eprint{1012.6039},
  \urlprefix\url{https://arxiv.org/abs/1012.6039}.

\bibitem[{\citenamefont{Philcox}(2023)}]{PhilcoxNpoint}
\bibinfo{author}{\bibfnamefont{O.~H.~E.} \bibnamefont{Philcox}}
  (\bibinfo{year}{2023}), \eprint{2303.08828},
  \urlprefix\url{https://arxiv.org/abs/2303.08828}.

\bibitem[{\citenamefont{{Hivon} et~al.}(2002)\citenamefont{{Hivon},
  {G{\'o}rski}, {Netterfield}, {Crill}, {Prunet}, and {Hansen}}}]{Hivon2002}
\bibinfo{author}{\bibfnamefont{E.}~\bibnamefont{{Hivon}}},
  \bibinfo{author}{\bibfnamefont{K.~M.} \bibnamefont{{G{\'o}rski}}},
  \bibinfo{author}{\bibfnamefont{C.~B.} \bibnamefont{{Netterfield}}},
  \bibinfo{author}{\bibfnamefont{B.~P.} \bibnamefont{{Crill}}},
  \bibinfo{author}{\bibfnamefont{S.}~\bibnamefont{{Prunet}}}, \bibnamefont{and}
  \bibinfo{author}{\bibfnamefont{F.}~\bibnamefont{{Hansen}}},
  \bibinfo{journal}{\apj} \textbf{\bibinfo{volume}{567}}, \bibinfo{pages}{2}
  (\bibinfo{year}{2002}), \eprint{astro-ph/0105302},
  \urlprefix\url{https://arxiv.org/abs/astro-ph/0105302}.

\bibitem[{\citenamefont{Surrao et~al.}(2023)\citenamefont{Surrao, Philcox, and
  Hill}}]{remastered}
\bibinfo{author}{\bibfnamefont{K.~M.} \bibnamefont{Surrao}},
  \bibinfo{author}{\bibfnamefont{O.~H.~E.} \bibnamefont{Philcox}},
  \bibnamefont{and} \bibinfo{author}{\bibfnamefont{J.~C.} \bibnamefont{Hill}},
  \bibinfo{journal}{Phys. Rev. D} \textbf{\bibinfo{volume}{107}},
  \bibinfo{pages}{083521} (\bibinfo{year}{2023}), \eprint{2302.05436},
  \urlprefix\url{https://arxiv.org/abs/2302.05436}.

\bibitem[{\citenamefont{{Zeldovich} and {Sunyaev}}(1969)}]{SZ1969}
\bibinfo{author}{\bibfnamefont{Y.~B.} \bibnamefont{{Zeldovich}}}
  \bibnamefont{and} \bibinfo{author}{\bibfnamefont{R.~A.}
  \bibnamefont{{Sunyaev}}}, \bibinfo{journal}{Astrophysics and Space Science}
  \textbf{\bibinfo{volume}{4}}, \bibinfo{pages}{301} (\bibinfo{year}{1969}).

\bibitem[{\citenamefont{{Sunyaev} and {Zeldovich}}(1970)}]{SZ1970}
\bibinfo{author}{\bibfnamefont{R.~A.} \bibnamefont{{Sunyaev}}}
  \bibnamefont{and} \bibinfo{author}{\bibfnamefont{Y.~B.}
  \bibnamefont{{Zeldovich}}}, \bibinfo{journal}{Astrophysics and Space Science}
  \textbf{\bibinfo{volume}{7}}, \bibinfo{pages}{3} (\bibinfo{year}{1970}).

\bibitem[{\citenamefont{Birkinshaw}(1999)}]{Birkinshaw_1999}
\bibinfo{author}{\bibfnamefont{M.}~\bibnamefont{Birkinshaw}},
  \bibinfo{journal}{Physics Reports} \textbf{\bibinfo{volume}{310}},
  \bibinfo{pages}{97} (\bibinfo{year}{1999}),
  \urlprefix\url{https://arxiv.org/abs/astro-ph/9808050}.

\bibitem[{\citenamefont{Stein et~al.}(2020)\citenamefont{Stein, Alvarez, Bond,
  van Engelen, and Battaglia}}]{Websky_2020}
\bibinfo{author}{\bibfnamefont{G.}~\bibnamefont{Stein}},
  \bibinfo{author}{\bibfnamefont{M.~A.} \bibnamefont{Alvarez}},
  \bibinfo{author}{\bibfnamefont{J.~R.} \bibnamefont{Bond}},
  \bibinfo{author}{\bibfnamefont{A.}~\bibnamefont{van Engelen}},
  \bibnamefont{and}
  \bibinfo{author}{\bibfnamefont{N.}~\bibnamefont{Battaglia}},
  \bibinfo{journal}{Journal of Cosmology and Astroparticle Physics}
  \textbf{\bibinfo{volume}{2020}}, \bibinfo{pages}{012} (\bibinfo{year}{2020}),
  \urlprefix\url{https://arxiv.org/abs/2001.08787}.

\bibitem[{\citenamefont{Tinker et~al.}(2008)\citenamefont{Tinker, Kravtsov,
  Klypin, Abazajian, Warren, Yepes, Gottlober, and Holz}}]{Tinker:2008}
\bibinfo{author}{\bibfnamefont{J.~L.} \bibnamefont{Tinker}},
  \bibinfo{author}{\bibfnamefont{A.~V.} \bibnamefont{Kravtsov}},
  \bibinfo{author}{\bibfnamefont{A.}~\bibnamefont{Klypin}},
  \bibinfo{author}{\bibfnamefont{K.}~\bibnamefont{Abazajian}},
  \bibinfo{author}{\bibfnamefont{M.~S.} \bibnamefont{Warren}},
  \bibinfo{author}{\bibfnamefont{G.}~\bibnamefont{Yepes}},
  \bibinfo{author}{\bibfnamefont{S.}~\bibnamefont{Gottlober}},
  \bibnamefont{and} \bibinfo{author}{\bibfnamefont{D.~E.} \bibnamefont{Holz}},
  \bibinfo{journal}{Astrophys. J.} \textbf{\bibinfo{volume}{688}},
  \bibinfo{pages}{709} (\bibinfo{year}{2008}), \eprint{0803.2706},
  \urlprefix\url{https://arxiv.org/pdf/0803.2706.pdf}.

\bibitem[{\citenamefont{Aghanim
  et~al.}(2020{\natexlab{b}})}]{Planck:2018cosmoparams}
\bibinfo{author}{\bibfnamefont{N.}~\bibnamefont{Aghanim}} \bibnamefont{et~al.}
  (\bibinfo{collaboration}{Planck}), \bibinfo{journal}{Astron. Astrophys.}
  \textbf{\bibinfo{volume}{641}}, \bibinfo{pages}{A6}
  (\bibinfo{year}{2020}{\natexlab{b}}), \bibinfo{note}{[Erratum:
  Astron.Astrophys. 652, C4 (2021)]}, \eprint{1807.06209},
  \urlprefix\url{https://arxiv.org/pdf/1807.06209.pdf}.

\bibitem[{\citenamefont{Battaglia et~al.}(2012)\citenamefont{Battaglia, Bond,
  Pfrommer, and Sievers}}]{Battaglia_2012}
\bibinfo{author}{\bibfnamefont{N.}~\bibnamefont{Battaglia}},
  \bibinfo{author}{\bibfnamefont{J.~R.} \bibnamefont{Bond}},
  \bibinfo{author}{\bibfnamefont{C.}~\bibnamefont{Pfrommer}}, \bibnamefont{and}
  \bibinfo{author}{\bibfnamefont{J.~L.} \bibnamefont{Sievers}},
  \bibinfo{journal}{The Astrophysical Journal} \textbf{\bibinfo{volume}{758}},
  \bibinfo{pages}{75} (\bibinfo{year}{2012}),
  \urlprefix\url{https://arxiv.org/abs/1109.3711}.

\bibitem[{\citenamefont{{Knox}}(1995)}]{Knox1995}
\bibinfo{author}{\bibfnamefont{L.}~\bibnamefont{{Knox}}},
  \bibinfo{journal}{\prd} \textbf{\bibinfo{volume}{52}}, \bibinfo{pages}{4307}
  (\bibinfo{year}{1995}), \eprint{astro-ph/9504054},
  \urlprefix\url{https://arxiv.org/abs/astro-ph/9504054}.

\bibitem[{\citenamefont{Surrao and Hill}(to appear)}]{PaperII}
\bibinfo{author}{\bibfnamefont{K.~M.} \bibnamefont{Surrao}} \bibnamefont{and}
  \bibinfo{author}{\bibfnamefont{J.~C.} \bibnamefont{Hill}} (\bibinfo{year}{to
  appear}).

\bibitem[{\citenamefont{{Lueckmann} et~al.}(2017)\citenamefont{{Lueckmann},
  {Goncalves}, {Bassetto}, {{\"O}cal}, {Nonnenmacher}, and {Macke}}}]{SNPE}
\bibinfo{author}{\bibfnamefont{J.-M.} \bibnamefont{{Lueckmann}}},
  \bibinfo{author}{\bibfnamefont{P.~J.} \bibnamefont{{Goncalves}}},
  \bibinfo{author}{\bibfnamefont{G.}~\bibnamefont{{Bassetto}}},
  \bibinfo{author}{\bibfnamefont{K.}~\bibnamefont{{{\"O}cal}}},
  \bibinfo{author}{\bibfnamefont{M.}~\bibnamefont{{Nonnenmacher}}},
  \bibnamefont{and} \bibinfo{author}{\bibfnamefont{J.~H.}
  \bibnamefont{{Macke}}}, \bibinfo{journal}{arXiv e-prints}
  \bibinfo{eid}{arXiv:1711.01861} (\bibinfo{year}{2017}), \eprint{1711.01861},
  \urlprefix\url{https://arxiv.org/pdf/1711.01861.pdf}.

\bibitem[{\citenamefont{{G{\'o}rski} et~al.}(2005)\citenamefont{{G{\'o}rski},
  {Hivon}, {Banday}, {Wandelt}, {Hansen}, {Reinecke}, and
  {Bartelmann}}}]{Healpix}
\bibinfo{author}{\bibfnamefont{K.~M.} \bibnamefont{{G{\'o}rski}}},
  \bibinfo{author}{\bibfnamefont{E.}~\bibnamefont{{Hivon}}},
  \bibinfo{author}{\bibfnamefont{A.~J.} \bibnamefont{{Banday}}},
  \bibinfo{author}{\bibfnamefont{B.~D.} \bibnamefont{{Wandelt}}},
  \bibinfo{author}{\bibfnamefont{F.~K.} \bibnamefont{{Hansen}}},
  \bibinfo{author}{\bibfnamefont{M.}~\bibnamefont{{Reinecke}}},
  \bibnamefont{and}
  \bibinfo{author}{\bibfnamefont{M.}~\bibnamefont{{Bartelmann}}},
  \bibinfo{journal}{\apj} \textbf{\bibinfo{volume}{622}}, \bibinfo{pages}{759}
  (\bibinfo{year}{2005}), \eprint{astro-ph/0409513},
  \urlprefix\url{https://arxiv.org/abs/astro-ph/0409513}.

\bibitem[{\citenamefont{Zonca et~al.}(2019)\citenamefont{Zonca, Singer, Lenz,
  Reinecke, Rosset, Hivon, and Gorski}}]{Healpy}
\bibinfo{author}{\bibfnamefont{A.}~\bibnamefont{Zonca}},
  \bibinfo{author}{\bibfnamefont{L.}~\bibnamefont{Singer}},
  \bibinfo{author}{\bibfnamefont{D.}~\bibnamefont{Lenz}},
  \bibinfo{author}{\bibfnamefont{M.}~\bibnamefont{Reinecke}},
  \bibinfo{author}{\bibfnamefont{C.}~\bibnamefont{Rosset}},
  \bibinfo{author}{\bibfnamefont{E.}~\bibnamefont{Hivon}}, \bibnamefont{and}
  \bibinfo{author}{\bibfnamefont{K.}~\bibnamefont{Gorski}},
  \bibinfo{journal}{Journal of Open Source Software}
  \textbf{\bibinfo{volume}{4}}, \bibinfo{pages}{1298} (\bibinfo{year}{2019}),
  \urlprefix\url{https://doi.org/10.21105/joss.01298}.

\bibitem[{\citenamefont{Harris et~al.}(2020)\citenamefont{Harris, Millman,
  van~der Walt, Gommers, Virtanen, Cournapeau, Wieser, Taylor, Berg, Smith
  et~al.}}]{numpy}
\bibinfo{author}{\bibfnamefont{C.~R.} \bibnamefont{Harris}},
  \bibinfo{author}{\bibfnamefont{K.~J.} \bibnamefont{Millman}},
  \bibinfo{author}{\bibfnamefont{S.~J.} \bibnamefont{van~der Walt}},
  \bibinfo{author}{\bibfnamefont{R.}~\bibnamefont{Gommers}},
  \bibinfo{author}{\bibfnamefont{P.}~\bibnamefont{Virtanen}},
  \bibinfo{author}{\bibfnamefont{D.}~\bibnamefont{Cournapeau}},
  \bibinfo{author}{\bibfnamefont{E.}~\bibnamefont{Wieser}},
  \bibinfo{author}{\bibfnamefont{J.}~\bibnamefont{Taylor}},
  \bibinfo{author}{\bibfnamefont{S.}~\bibnamefont{Berg}},
  \bibinfo{author}{\bibfnamefont{N.~J.} \bibnamefont{Smith}},
  \bibnamefont{et~al.}, \bibinfo{journal}{Nature}
  \textbf{\bibinfo{volume}{585}}, \bibinfo{pages}{357} (\bibinfo{year}{2020}),
  \urlprefix\url{https://arxiv.org/pdf/2006.10256.pdf}.

\bibitem[{\citenamefont{Virtanen et~al.}(2020)\citenamefont{Virtanen, Gommers,
  Oliphant, Haberland, Reddy, Cournapeau, Burovski, Peterson, Weckesser, Bright
  et~al.}}]{scipy}
\bibinfo{author}{\bibfnamefont{P.}~\bibnamefont{Virtanen}},
  \bibinfo{author}{\bibfnamefont{R.}~\bibnamefont{Gommers}},
  \bibinfo{author}{\bibfnamefont{T.~E.} \bibnamefont{Oliphant}},
  \bibinfo{author}{\bibfnamefont{M.}~\bibnamefont{Haberland}},
  \bibinfo{author}{\bibfnamefont{T.}~\bibnamefont{Reddy}},
  \bibinfo{author}{\bibfnamefont{D.}~\bibnamefont{Cournapeau}},
  \bibinfo{author}{\bibfnamefont{E.}~\bibnamefont{Burovski}},
  \bibinfo{author}{\bibfnamefont{P.}~\bibnamefont{Peterson}},
  \bibinfo{author}{\bibfnamefont{W.}~\bibnamefont{Weckesser}},
  \bibinfo{author}{\bibfnamefont{J.}~\bibnamefont{Bright}},
  \bibnamefont{et~al.}, \bibinfo{journal}{Nature Methods}
  \textbf{\bibinfo{volume}{17}}, \bibinfo{pages}{261} (\bibinfo{year}{2020}),
  \urlprefix\url{https://arxiv.org/pdf/1907.10121.pdf}.

\bibitem[{\citenamefont{Hunter}(2007)}]{matplotlib}
\bibinfo{author}{\bibfnamefont{J.~D.} \bibnamefont{Hunter}},
  \bibinfo{journal}{Computing in Science \& Engineering}
  \textbf{\bibinfo{volume}{9}}, \bibinfo{pages}{90} (\bibinfo{year}{2007}),
  \urlprefix\url{https://ieeexplore.ieee.org/document/4160265}.

\bibitem[{\citenamefont{{Astropy Collaboration}
  et~al.}(2013)\citenamefont{{Astropy Collaboration}, {Robitaille}, {Tollerud},
  {Greenfield}, {Droettboom}, {Bray}, {Aldcroft}, {Davis}, {Ginsburg},
  {Price-Whelan} et~al.}}]{astropy1}
\bibinfo{author}{\bibnamefont{{Astropy Collaboration}}},
  \bibinfo{author}{\bibfnamefont{T.~P.} \bibnamefont{{Robitaille}}},
  \bibinfo{author}{\bibfnamefont{E.~J.} \bibnamefont{{Tollerud}}},
  \bibinfo{author}{\bibfnamefont{P.}~\bibnamefont{{Greenfield}}},
  \bibinfo{author}{\bibfnamefont{M.}~\bibnamefont{{Droettboom}}},
  \bibinfo{author}{\bibfnamefont{E.}~\bibnamefont{{Bray}}},
  \bibinfo{author}{\bibfnamefont{T.}~\bibnamefont{{Aldcroft}}},
  \bibinfo{author}{\bibfnamefont{M.}~\bibnamefont{{Davis}}},
  \bibinfo{author}{\bibfnamefont{A.}~\bibnamefont{{Ginsburg}}},
  \bibinfo{author}{\bibfnamefont{A.~M.} \bibnamefont{{Price-Whelan}}},
  \bibnamefont{et~al.}, \bibinfo{journal}{\aap} \textbf{\bibinfo{volume}{558}},
  \bibinfo{eid}{A33} (\bibinfo{year}{2013}), \eprint{1307.6212},
  \urlprefix\url{https://arxiv.org/pdf/1307.6212.pdf}.

\bibitem[{\citenamefont{{Astropy Collaboration}
  et~al.}(2018)\citenamefont{{Astropy Collaboration}, {Price-Whelan},
  {Sip{\H{o}}cz}, {G{\"u}nther}, {Lim}, {Crawford}, {Conseil}, {Shupe},
  {Craig}, {Dencheva} et~al.}}]{astropy2}
\bibinfo{author}{\bibnamefont{{Astropy Collaboration}}},
  \bibinfo{author}{\bibfnamefont{A.~M.} \bibnamefont{{Price-Whelan}}},
  \bibinfo{author}{\bibfnamefont{B.~M.} \bibnamefont{{Sip{\H{o}}cz}}},
  \bibinfo{author}{\bibfnamefont{H.~M.} \bibnamefont{{G{\"u}nther}}},
  \bibinfo{author}{\bibfnamefont{P.~L.} \bibnamefont{{Lim}}},
  \bibinfo{author}{\bibfnamefont{S.~M.} \bibnamefont{{Crawford}}},
  \bibinfo{author}{\bibfnamefont{S.}~\bibnamefont{{Conseil}}},
  \bibinfo{author}{\bibfnamefont{D.~L.} \bibnamefont{{Shupe}}},
  \bibinfo{author}{\bibfnamefont{M.~W.} \bibnamefont{{Craig}}},
  \bibinfo{author}{\bibfnamefont{N.}~\bibnamefont{{Dencheva}}},
  \bibnamefont{et~al.}, \bibinfo{journal}{\aj} \textbf{\bibinfo{volume}{156}},
  \bibinfo{eid}{123} (\bibinfo{year}{2018}), \eprint{1801.02634},
  \urlprefix\url{https://arxiv.org/pdf/1801.02634.pdf}.

\bibitem[{\citenamefont{{Astropy Collaboration}
  et~al.}(2022)\citenamefont{{Astropy Collaboration}, {Price-Whelan}, {Lim},
  {Earl}, {Starkman}, {Bradley}, {Shupe}, {Patil}, {Corrales}, {Brasseur}
  et~al.}}]{astropy3}
\bibinfo{author}{\bibnamefont{{Astropy Collaboration}}},
  \bibinfo{author}{\bibfnamefont{A.~M.} \bibnamefont{{Price-Whelan}}},
  \bibinfo{author}{\bibfnamefont{P.~L.} \bibnamefont{{Lim}}},
  \bibinfo{author}{\bibfnamefont{N.}~\bibnamefont{{Earl}}},
  \bibinfo{author}{\bibfnamefont{N.}~\bibnamefont{{Starkman}}},
  \bibinfo{author}{\bibfnamefont{L.}~\bibnamefont{{Bradley}}},
  \bibinfo{author}{\bibfnamefont{D.~L.} \bibnamefont{{Shupe}}},
  \bibinfo{author}{\bibfnamefont{A.~A.} \bibnamefont{{Patil}}},
  \bibinfo{author}{\bibfnamefont{L.}~\bibnamefont{{Corrales}}},
  \bibinfo{author}{\bibfnamefont{C.~E.} \bibnamefont{{Brasseur}}},
  \bibnamefont{et~al.}, \bibinfo{journal}{apj} \textbf{\bibinfo{volume}{935}},
  \bibinfo{eid}{167} (\bibinfo{year}{2022}), \eprint{2206.14220},
  \urlprefix\url{https://arxiv.org/pdf/2206.14220.pdf}.

\bibitem[{\citenamefont{Johansson and Forssen}(2016)}]{pywigxjpf}
\bibinfo{author}{\bibfnamefont{H.~T.} \bibnamefont{Johansson}}
  \bibnamefont{and} \bibinfo{author}{\bibfnamefont{C.}~\bibnamefont{Forssen}},
  \bibinfo{journal}{SIAM Journal on Scientific Computing}
  \textbf{\bibinfo{volume}{38}}, \bibinfo{pages}{A376} (\bibinfo{year}{2016}),
  \eprint{http://dx.doi.org/10.1137/15M1021908},
  \urlprefix\url{https://arxiv.org/pdf/1504.08329.pdf}.

\end{thebibliography}
\end{document}